\documentclass[aps,prl,reprint,amsmath,amssymb,superscriptaddress,nofootinbib,noeprint]{revtex4-2}
\pdfoutput=1

\usepackage{soul}
\usepackage{graphicx}% Include figure files
\usepackage{dcolumn}% Align table columns on decimal point
\usepackage{bm}% bold math
\usepackage[version=3]{mhchem}
\usepackage{natbib}
\usepackage{mathtools}
\PassOptionsToPackage{hyphens}{url}
\usepackage[colorlinks=true,allcolors=blue]{hyperref}
\usepackage{lineno}
\usepackage{float}
\usepackage{color}
\usepackage{graphicx}
\usepackage{amsmath,amssymb}
\usepackage{upgreek}
\usepackage{psfrag} 
\usepackage{graphicx}
\usepackage{braket}
\usepackage{units}
\usepackage{times}
\usepackage[svgnames]{xcolor}
\definecolor{myColor}{rgb}{0.02,0.12,0.3}
\definecolor{myciteColor}{rgb}{0.39,0.7,0.89}
\usepackage[colorlinks=true,allcolors=blue]{hyperref}
\newcommand{\F}{\mathrm{F}}
\newcommand{\muw}{{\mu\mathrm{w}}}
\newcommand{\B}{\mathrm{B}}

\newcommand{\dd }{\mathrm{d}}

\newcommand{\TTFi}{(T/T_\F)_\mathrm{i}}
\makeatletter
\def\maketitle{
\@author@finish
\title@column\titleblock@produce
\suppressfloats[t]}
\begin{document}
%\linenumbers
\title{%Quantum Joule-Thomson Effect in Universal Fermi Gases\\
Observation of the Fermionic Joule-Thomson Effect}

\author{Yunpeng Ji}
\email[To whom correspondence should be addressed:\\ yunpeng.ji@yale.edu.]{}
\affiliation{Department of Physics, Yale University, New Haven, Connecticut 06520, USA}
\author{Jianyi Chen}
\affiliation{Department of Physics, Yale University, New Haven, Connecticut 06520, USA}
\author{Grant L. Schumacher}
\affiliation{Department of Physics, Yale University, New Haven, Connecticut 06520, USA}
\author{Gabriel G. T. Assump{\c{c}}{\~a}o}
\affiliation{Department of Physics, Yale University, New Haven, Connecticut 06520, USA}
\author{Songtao Huang}
\affiliation{Department of Physics, Yale University, New Haven, Connecticut 06520, USA}
\author{Franklin J. Vivanco}
\affiliation{Department of Physics, Yale University, New Haven, Connecticut 06520, USA}
\author{Nir Navon}
\affiliation{Department of Physics, Yale University, New Haven, Connecticut 06520, USA}
\affiliation{Yale Quantum Institute, Yale University, New Haven, Connecticut 06520, USA}

\date{\today}

\begin{abstract}
We report the observation of the quantum Joule-Thomson (JT) effect in ideal and unitary Fermi gases. We study the temperature dynamics of these systems while they undergo an energy-per-particle conserving rarefaction. For scale-invariant systems, whose equations of state satisfy the relation $U\propto PV$, this rarefaction conserves the specific enthalpy, which makes it thermodynamically equivalent to a JT throttling process. We observe JT heating in an ideal Fermi gas, stronger at higher quantum degeneracy, a result of the repulsive quantum-statistical `force' arising from Pauli blocking. In a unitary Fermi gas, we observe that the JT heating is marginal in the temperature range $0.2 \lesssim T/T_{\F} \lesssim 0.8 $ as the repulsive quantum-statistical effect is lessened by the attractive interparticle interaction.
\end{abstract}

\maketitle
The Joule-Thomson (JT) effect is a fundamental phenomenon in thermodynamics whereby the temperature $T$ of a thermally isolated system changes in response to a change (typically a decrease) of the pressure $P$ while the specific enthalpy $h$ is conserved. This effect has played a momentous role in the history of thermodynamics~\cite{joule_1852} and the birth of modern cryogenics~\cite{linde_1899}. In its own right, the JT effect has attracted interest as a probe of the thermodynamics of imperfect (\emph{i.e.} interacting) gases~\cite{roebuck_1926,hirschfelder_1938} and, more recently, in relation to black hole expansion dynamics~\cite{Okcu_2017,ghaffarnejad_2018,mo_2018}.

In classical gases, the JT effect is tied to interparticle interactions. This can be illustrated with a simple equation of state (EoS) $PV=Nk_{\B}T + a_{\mathrm{int}} N^2/V$, \emph{i.e.} the van der Waals EoS without the excluded-volume effect; $N$ is the number of particles, $V$ is the volume, and $a_{\mathrm{int}}$ is an interaction parameter that is positive for repulsive interactions and negative for attractive ones. In the limit of weak interactions ($a_{\mathrm{int}} N/V \ll k_{\B}T$), the Joule-Thomson coefficient $\mu_{\mathrm{JT}}\equiv\left(\partial T/\partial P\right)_h$ is $\mu_{\mathrm{JT}}\propto-a_{\mathrm{int}}/c_P$ in this model (where $c_P$ is the specific heat); thus the system heats (resp. cools) in the case of repulsive (resp. attractive) interactions.

 Surprisingly, the JT effect does not require interactions. Indeed, shortly after the discovery of quantum indistinguishability, it was predicted that quantum correlations give rise to a nontrivial JT effect even in the absence of interactions~\cite{kothari_1937}; in essence, Fermi-Dirac particles would behave as if they were classically repelling (and Bose-Einstein particles, as if they were attracting). Despite their fundamental nature, the bosonic JT effect was only recently observed~\cite{Schmidutz_2014} - enabled by the creation of homogeneous Bose gases - whereas the fermionic one has remained elusive.

In this work we measure the JT effect in Fermi systems. In the textbook presentation of the JT process, a gas is throttled through a porous plug from a high-$P$ to a low-$P$ compartment (see sketch in Fig.~\ref{Fig:1}(a)). In our experiment, we realize a JT rarefaction either by exploiting collisions with high-energy particles from the residual background gas (in the vacuum chamber) or by controllably transferring atoms into internal states that are essentially not interacting with the states of interest (Fig.~\ref{Fig:1}(b)). In either case, the loss process is independent of the energy per particle $u = U/N$ where $U$ is the total internal energy. For a scale-invariant gas, whose EoS satisfies $U\propto PV$, this process is thermodynamically equivalent to a JT one.

\begin{figure*}[!hbt]
\includegraphics[width=2\columnwidth]{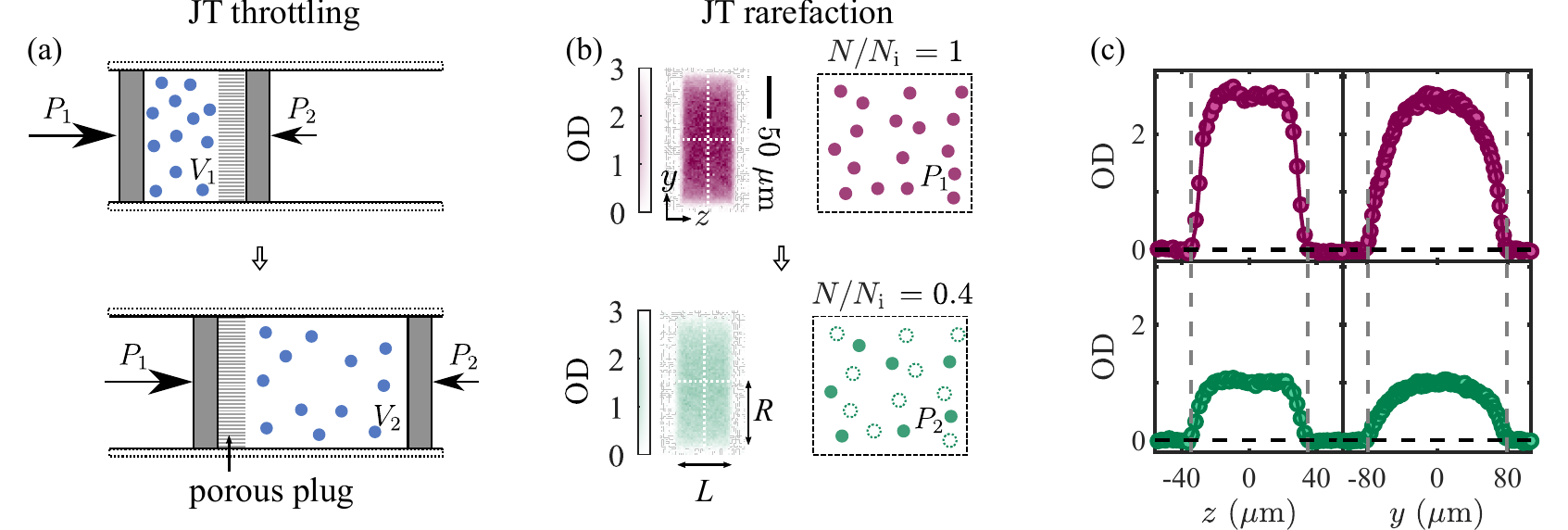}
\caption{Joule-Thomson rarefaction of a homogeneous Fermi gas. (a) Classical throttling process. A thermally isolated gas is forced from a high pressure chamber (top) to a low pressure one (bottom). (b) JT process through a rarefaction at fixed energy per particle $u$ and volume $V$. The images to the left of the cartoons are \emph{in situ} optical density (OD) images of homogeneous Fermi gases of $^6$Li atoms, before (top, purple) and after (bottom, green) rarefaction. (c) Cuts along the white dashed lines on the OD images. The solid lines are fits to extract the volume of the box. The length and radius of this cylindrical box are $L=58(1)$~$\mu\mathrm{m}$ and $R=77(2)$~$\mu\mathrm{m}$, and remain essentially unchanged during rarefaction (see vertical dashed grey lines as guides to the eye). The density cuts are colored according to panel (b).}
\label{Fig:1}
\end{figure*}

\begin{figure}[!hbt]
\includegraphics[width=1\columnwidth]{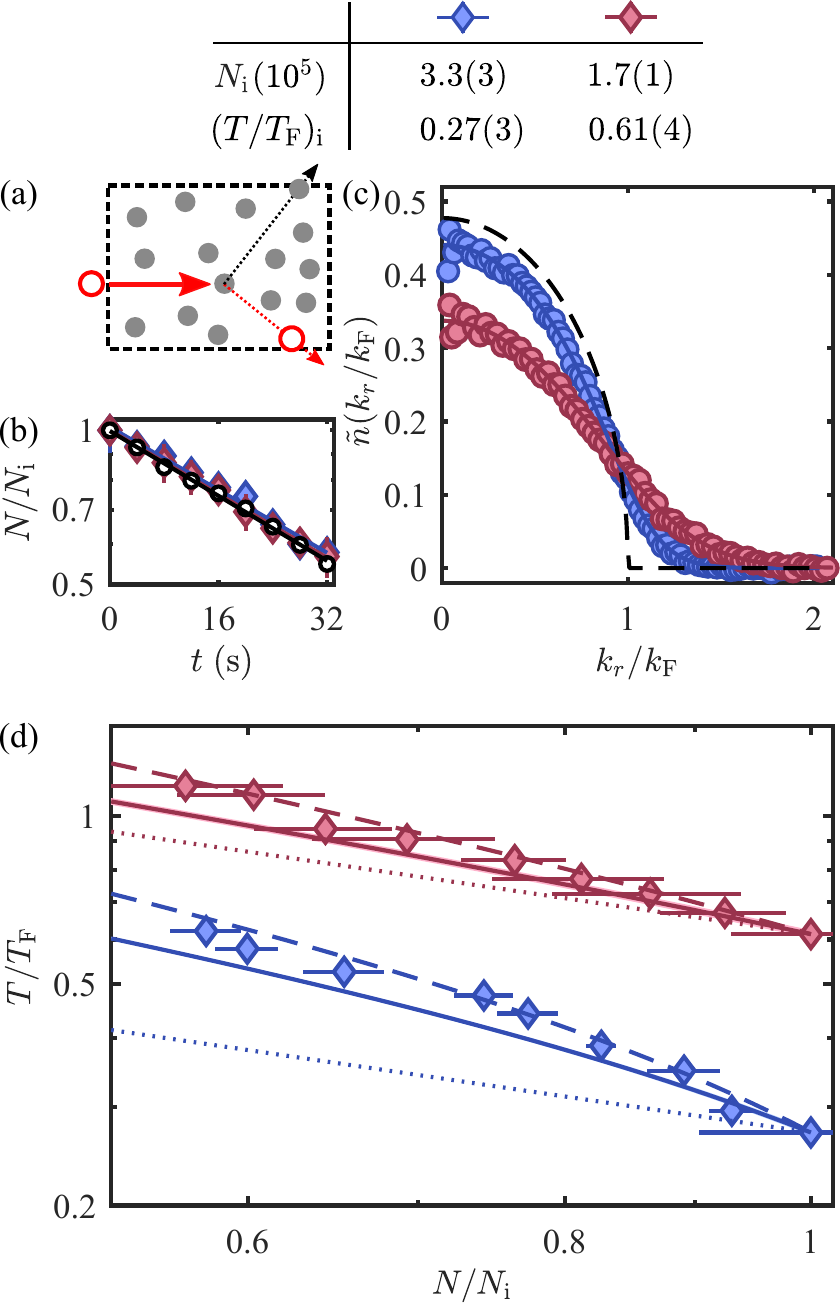}
\caption{Joule-Thomson effect of the ideal Fermi gas. (a) Sketch of energy-independent atom loss due to collisions with high-energy background particles (red empty circle). (b) Decay of a non-interacting Fermi gas (black circles) and weakly interacting Fermi gases (colored diamonds) at different initial $T/T_{\F}$ (see legend). The solid lines are exponential fits. The same marker color is used in (c) and (d). (c) Azimuthally-averaged radial momentum distribution of Fermi gases extracted from column integrated OD after a time-of-flight expansion of duration $t_\text{TOF}$. The distributions are normalized such that $\int \tilde n(k_r/k_\F) (2\pi k_r/k_\F) \dd (k_r/k_\F) = 1$; here $k_r=m\sqrt{y^2+z^2}/(\hbar t_\text{TOF})$ and $k_\F$ is the Fermi wavenumber. The distributions correspond to the initial points in (b). The solid lines are fits to Fermi-Dirac distributions to extract temperatures. The black dashed line shows the momentum distribution at $T = 0$. (d) Temperature evolution of weakly interacting Fermi gases during a JT rarefaction. The solid lines are theoretical predictions fixing $\TTFi$ to the experimentally measured values, with the (barely visible) bands representing the uncertainty on $\TTFi$. The dashed lines take into account the effect of technical heating in the box~\cite{SuppMat}. The dotted lines show the evolution of $T/T_{\F}$ at constant $T$.}
\label{Fig:2}
\end{figure}

We first focus on the JT effect in the ideal Fermi gas. We prepare weakly interacting spin-$1/2$ Fermi gases of $^6$Li atoms in a balanced mixture of the first and third lowest Zeeman sublevels (respectively labelled $\ket{1}$ and $\ket{3}$). Our gases are confined in optical boxes so that their density and other thermodynamics quantities are spatially uniform, making the interpretation of our measurements straightforward~\cite{navon_2021}. Our cylindrical boxes have a radius $R=77(2)$~$\mu$m and an adjustable length $L$ between $58~\mu$m and $120~\mu$m (see an example in Fig.~\ref{Fig:1}(c)). The samples are evaporated at a bias magnetic field $B=287$~G, where the s-wave scattering length $a \approx -280 a_0$ ($a_0$ is the Bohr radius). Levitation against gravity is done with a magnetic field gradient. We typically start our experiments with a degenerate spin-1/2 Fermi gas, $T\lesssim E_\F/k_\B$ (where $E_\F=\hbar^2/(2m)(6\pi^2 N/V)^{2/3}$ is the Fermi energy); henceforth, all thermodynamic quantities (such $U$, $N$, etc.) are defined for each spin population. We typically have $N_1\approx N_3\approx 8\times10^5$, corresponding to a Fermi temperature of $T_\F=E_\F/k_\B\approx 300$~nK.

We take advantage of the slow one-body losses due to collisions with the background gas to realize $u$-constant rarefactions (Fig.~\ref{Fig:2}(a)), as in~\cite{Schmidutz_2014}. Here, the tunability of interparticle interactions is important as the interactions must obey conflicting requirements. On the one hand, interactions must be weak enough so that we probe essentially ideal gas physics and that $PV=(2/3)U$; furthermore, two-body energy-dependent evaporation must be suppressed on the (long) timescale of the measurements. On the other hand, interactions must be strong enough to ensure that the gas is in thermal equilibrium when the measurements are done. 

Consequently, we satisfy those conditions by choosing $a$ in the range $100 a_0 \lesssim a \lesssim 220 a_0$. The specific value is picked as large as possible, while ensuring that the decay time is indistinguishable from the vacuum-limited lifetime. In Fig.~\ref{Fig:2}(b), we show examples of decays at various quantum degeneracies (colored diamonds); each data set is normalized to the initial number of atoms $N_{\mathrm{i}}$. The lifetimes in those data series are indistinguishable from that of a non-interacting gas ($|a|\le 50 a_0$), $\tau_{\text{vac}} = ~55(2)$~s (see the black circles, the vacuum-limited lifetime in our chamber). At the same time, the two-body elastic collision rate is in the appropriate regime, $\Gamma_\mathrm{el} \gg 1/\tau_{\text{vac}}$ (in our range of densities and temperatures, $\Gamma_\mathrm{el} \geq 0.17~\mathrm{s}^{-1}$). Furthermore, for our regime of interactions and temperatures, $h$ changes less than $0.01\%$ during the decay, making this rarefaction an excellent approximation of a JT process~\cite{SuppMat}.

Thermometry is performed using time-of-flight expansions. In our case, the gas parameter $k_{\F} a \le 0.02$, so interaction effects in flight are weak, \emph{i.e.} the flights are essentially ballistic. Furthermore, $k_{\F} a$ is low enough so that interactions do not appreciably affect the \emph{in situ} momentum distribution~\cite{SuppMat}; thermometry can thus be done as if the samples were non-interacting (Fig.~\ref{Fig:2}(c)). 

 In Fig.~\ref{Fig:2}(d), we show the temperature dynamics of the gas during rarefaction for the initial conditions $\TTFi=0.27(3)$ (blue) and $\TTFi=0.61(4)$ (red). We plot $T/T_\F$, where the instantaneous $T_\F$ decreases as the gas rarefies. The dotted lines correspond to $T/T_\F \propto (N/N_\mathrm{i})^{-2/3}$, the expectation for constant-$T$ rarefactions. The measurements show heating, and the main qualitative feature is that the heating is more pronounced for a more quantum-degenerate gas.

Quantitatively, we describe the temperature dynamics during this JT process using the dimensionless coefficient $\theta_{\mathrm{JT}} \equiv (\partial \log(T)/\partial \log(P))_h$. This coefficient is related to the Joule-Thomson coefficient: $\mu_{\mathrm{JT}}=(T/P)\theta_{\mathrm{JT}}$. For a homogeneous gas whose EoS is universal, \emph{i.e.} for which $P \lambda_{T}^3/(k_{\B}T)$ only depends on the chemical potential $\mu$ and $k_\B T$ via the ratio $\mu/(k_\B T)$, $\theta_{\mathrm{JT}}$ is a function of $T/T_\F$ alone. The evolution of $T/T_\F$ follows $(\partial \log(T/T_\F)/\partial \log(N))_h = \theta_{\mathrm{JT}} - 2/3$~
\cite{SuppMat}. In Fig.~\ref{Fig:2}(d), solid lines are the theoretical predictions derived from the EoS of the ideal Fermi gas (where $\TTFi$ is fixed to the experimental value). We find good agreement with the data. The small discrepancy is well accounted for by a weak technical heating in our box; the dashed lines show the theoretical predictions from the model $\dd \log(T/T_\F)/\dd \log(N) = (\theta_{\mathrm{JT}} - 2/3)(1+(3/2)\gamma_{\text{tech}}\tau_{\text{vac}}/u)$, where our heating rate  $\gamma_{\text{tech}}=0.58(7)k_\B\times$nK/s is characterized independently~\cite{SuppMat}.

\begin{figure}[!hbt]
\includegraphics[width=1\columnwidth]{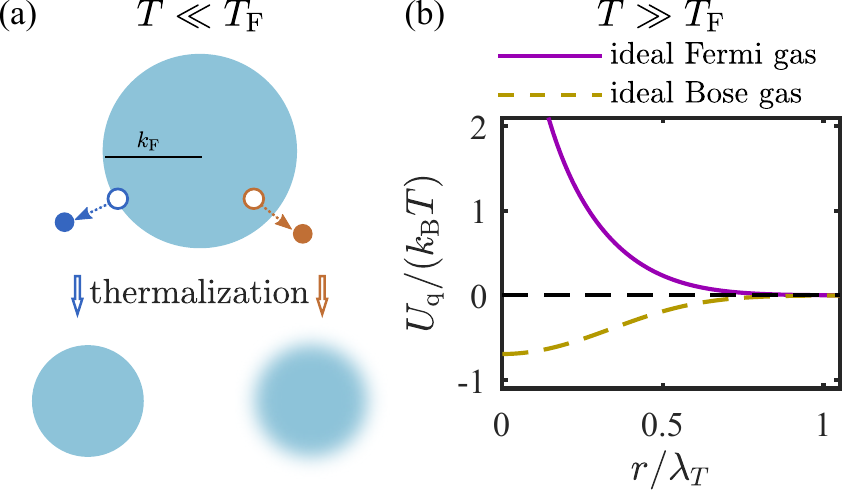}
\caption{Microscopic interpretations of the fermionic JT effect. (a) Sketch of Fermi hole heating at $T \ll T_\F$. The blue (resp. brown) point represents a particle whose removal causes no temperature change (resp. heating). (b) Quantum-statistical interaction potential of ideal quantum gases in the high-$T$ limit.}
\label{Fig:3}
\end{figure}
% and the unitary Fermi gas 
In the low- and high-$T$ limits, simple pictures provide insights into the microscopic origin of the JT effect. First, for $T\ll T_\F$, the state of the gas is essentially a Fermi sea (Fig.~\ref{Fig:3}(a)). In that case, the average energy per particle lost in a random (energy-independent) removal is only $u_{\mathrm{loss}} \approx (3/5) E_\F$; the energy per particle that needs to be removed to keep the temperature constant, $u_T\equiv \left(\partial U/\partial N\right)_{T,V}$, is $u_T\approx E_\F$. As a result, the system heats up, a process referred to as Fermi hole heating~\cite{Timmermans_2001}. 

The interpretation of quantum correlations as statistical `forces' demystifies the quantum JT effect in the $T\gg T_\F$ limit. For that purpose, it is useful to consider the pair density correlation function $G(\mathbf{r}_1,\mathbf{r}_2)\equiv \langle \Psi^\dagger(\mathbf{r}_1)\Psi^\dagger(\mathbf{r}_2) \Psi(\mathbf{r}_2)\Psi(\mathbf{r}_1)\rangle / (n(\mathbf{r}_1)n(\mathbf{r}_2))$, where $\Psi^\dagger(\mathbf{r}_j)$ ($\Psi(\mathbf{r}_j)$) is the field operator that creates (annihilates) a particle at position $\mathbf{r}_j$, and $n(\mathbf{r}_j)\equiv\langle \Psi^\dagger(\mathbf{r}_j)\Psi(\mathbf{r}_j)\rangle$.
For an ideal homogeneous gas in the high-$T$ (virial) limit, $G(\mathbf{r}_1,\mathbf{r}_2)=G(r)\approx 1 + \eta \exp(-2 \pi r^2/\lambda_{T}^2)$, where $r=|\mathbf{r}_1-\mathbf{r}_2|$, $\eta = 1$ for bosons and $\eta = -1$ for fermions ($\eta=0$ for the classical ideal gas), and $\lambda_T$ is the thermal wavelength~\cite{pathria_2011}. For a dilute classical gas, $G(r) \approx \exp(-U_\mathrm{int}(r)/(k_{\B} T))$, where $U_\mathrm{int}(r)$ is the interparticle interaction potential. By analogy, one can define an effective quantum-statistical interaction between indistinguishable non-interacting particles, $U_\mathrm{q}(r)\equiv -k_\B T\log G(r)$~\cite{uhlenbeck_1932,mullin_2003}. The potential $U_\mathrm{q}(r)$ is shown as yellow and purple lines in Fig.~\ref{Fig:3}(b); as intuitively expected, fermions effectively `repel' while bosons `attract' each other. Furthermore, the sign of their quantum JT effect is consistent with their respective quantum-statistical interaction~\cite{virial}. 

We now turn to the unitary Fermi gas, for which $1/a=0$. Crucially, because $PV-(2/3)U \propto \mathcal{I}/a$, where $\mathcal{I}$ is Tan's contact~\cite{Tan_2008_Virial}, the universal relation $PV=(2/3)U$ also holds for the unitary gas; this makes the unitary case another special point in the BEC-BCS crossover~\cite{zwerger_2011} for which a $u$-constant rarefaction is also a JT process. 
We create a unitary gas by preparing a spin-balanced mixture of atoms in states $\ket{1}$ and $\ket{3}$ that is evaporatively cooled and loaded into the optical box at $B\approx796$~G. The field is then ramped to the Feshbach resonance, $B\approx 690$~G. At this stage we typically have $N_1\approx N_3 \approx 3\times10^5$ at $T/T_\F\approx0.2$ (slightly above the superfluid transition temperature $T_\mathrm{c}$~\cite{zwerger_2011}).

Just as in the ideal gas case, the two main ingredients to observe the JT effect in this setting are the realization of a $u$-constant rarefaction and a thermometry method. Both present new challenges compared to the weakly interacting case.

As the collision rate in the unitary gas is so high (typically $\Gamma_{\mathrm{el}}^\mathrm{uni} \ge 500\;\mathrm{s}^{-1}$ in our case), the evaporation rate is vastly higher than in the weakly interacting case, threatening the JT nature of the rarefaction. In our deepest box ($U_{\text{box}} \gtrsim 8E_\F$), the lifetime of our unitary gas is $\tau_\mathrm{uni}\approx 30~$s, close but a little shorter than $\tau_{\text{vac}}$~\cite{unitarylifetime} (possibly limited by a slow residual evaporation). To mitigate this issue, we artificially increase the $u$-independent `loss' rate by applying a weak two-tone microwave pulse of duration $t_{\mu\mathrm{w}}$ to transfer atoms to higher Zeeman sublevels ($\ket{1}$ to $\ket{6}$, and $\ket{3}$ to $\ket{4}$, where $\ket{j}$ with $j=1,...,6$ are labelled from the ground up, see Fig.~\ref{Fig:4}(a))~\cite{microwave}. The power of the tones are adjusted so that the transfer rates on the two transitions are the same. Measuring the number of atoms remaining in $\ket{1}$ and $\ket{3}$, we find exponential decays with respective characteristic times
$\tau_{\mu\mathrm{w}} = 0.33(1)$~s and  $\tau_{\mu\mathrm{w}} = 0.35(1)$~s (pink and blue diamonds in Fig.~\ref{Fig:4}(b)). This time scale is such that $\tau_\mathrm{uni}\gg \tau_{\mu\mathrm{w}} \gg 1/\Gamma_{\mathrm{el}}^\mathrm{uni}$, \emph{i.e.} the microwave-induced rarefaction is slow compared to the elastic collision rate so that the gas remains in thermal equilibrium, but fast enough so that energy-dependent losses are negligible. We validate our microwave-induced rarefaction method on the now-verified case of the weakly interacting gas; the effect of the technical heating is now negligible because the timescale of the microwave-induced rarefaction is short, see~\cite{SuppMat}.

\begin{figure}[h!]
\includegraphics[width=1\columnwidth]{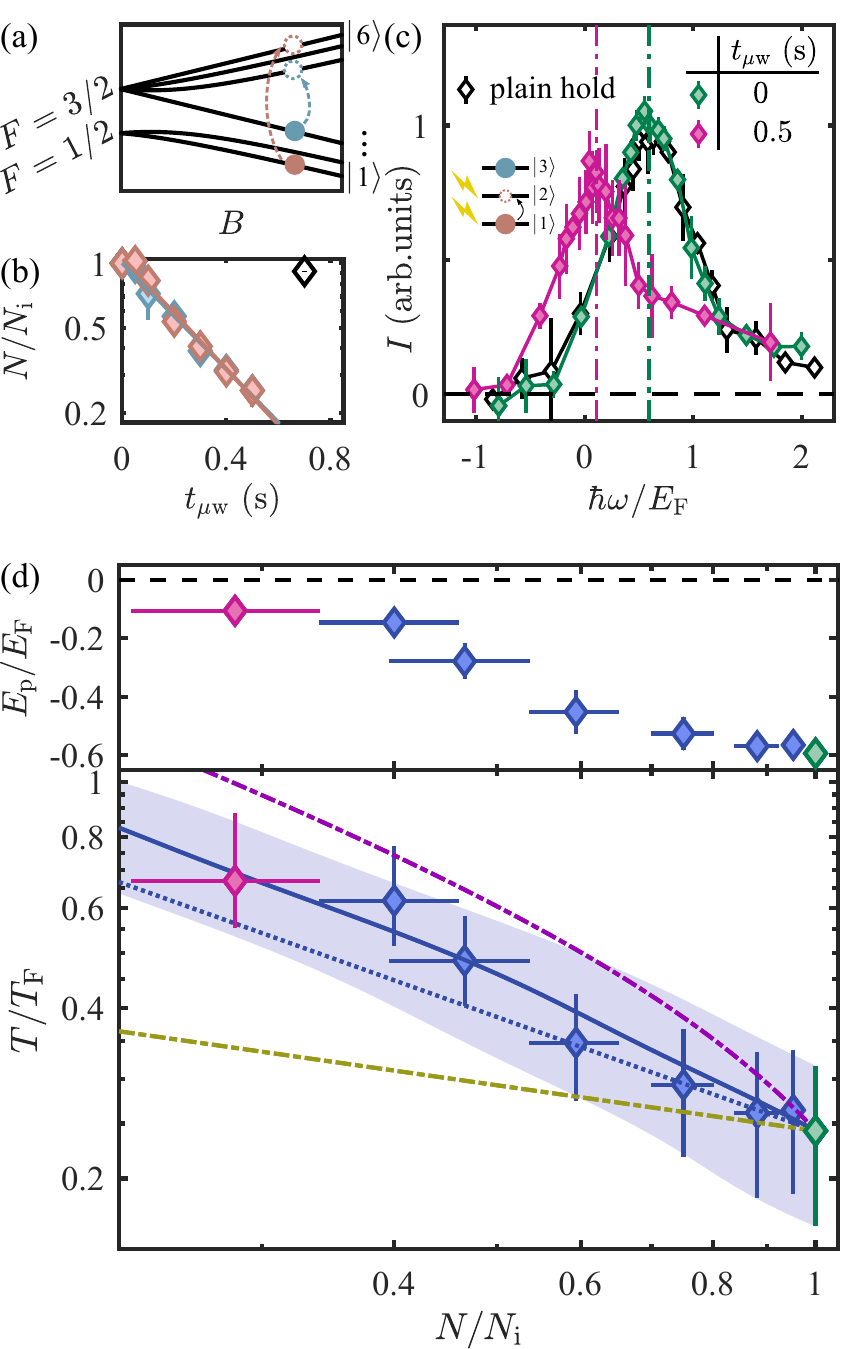}
\caption{Joule-Thomson effect of the unitary Fermi gas. (a) Breit-Rabi diagram of the ground state manifold of $^6\mathrm{Li}$ (sketch not to scale). Solid (open) symbols represent the initial (final) states of the microwave transfer. (b) Microwave-driven decay of a unitary Fermi gas. Pink and blue diamonds are the populations in state $\ket{1}$ and $\ket{3}$ respectively. The black empty diamond shows the reference decay without microwave field. (c) Radio-frequency (rf) thermometry. The cartoon shows the internal states used in the rf spectroscopy: the gas, initially in a balanced mixture of $\ket{1}$-$\ket{3}$, is driven on the transition $\ket{1}\rightarrow\ket{2}$; the states that are imaged are marked with the lightning symbols. Green and magenta diamonds are the spectra at $t_\muw = 0$~s and $t_\muw = 0.5$~s, and black diamonds correspond to the spectrum after $0.7$~s no-microwave hold. Dot-dashed vertical lines mark the peak response. (d) Degeneracy of a unitary Fermi gas during isenthalpic rarefaction. The peak frequency response $E_{\mathrm{p}}$ and $T/T_\F$ are shown along the rarefaction $N/N_\mathrm{i}$, respectively in the upper and lower panel. The green and magenta diamonds correspond to the spectra in (c). The blue solid line is the prediction based on the EoS~\cite{Ku_2012}, fixing $\TTFi$ to the experimental values. The blue band is the uncertainty arising from the uncertainty on $\TTFi$. The dotted line shows the evolution of $T/T_{\F}$ at constant $T$. The purple (resp. yellow) dot-dashed line is the theoretical temperature evolution of an ideal Fermi (resp. Bose) gas during JT process.}
\label{Fig:4}
\end{figure}

For thermometry, we use radio-frequency (rf) spectroscopy and compare it to the calibrated spectra for the unitary gas at finite temperature measured at MIT~\cite{mukherjee_2019,yan_2022}. We apply a $t_{\mathrm{pulse}} = 3$~ms square rf pulse with a (single-particle) Rabi frequency $\Omega_0 = 2\pi \times 139(1)$~Hz to transfer a small fraction of atoms ($\leq 15\%$) from state $\ket{1}$ to state $\ket{2}$. 
The normalized response spectrum, $I(\hbar \omega/ E_\F) = (N_2/N_1)E_{\F}/(\hbar\Omega_0^2 t_{\mathrm{pulse}})$ is temperature dependent ($N_1$ and $N_2$ are measured before and after the pulse respectively); here $\omega$ is measured relative to the bare $\ket{1}\rightarrow\ket{2}$ transition frequency, which is calibrated using a fully polarized sample prepared in state $\ket{1}$~\cite{SuppMat}. Specifically, we extract the temperature from the peak response frequency $E_{\mathrm{p}} \equiv -\hbar\omega_{\mathrm{p}}$, whose magnitude decreases monotonically with increasing $T/T_\F$~\cite{mukherjee_2019}.

We first verify that without microwave transfers and on the timescale of the experiment, evaporation and other $T$ dynamics are negligible. We measure the initial rf spectrum of the gas (green diamonds in Fig.\ref{Fig:4}(c)) and after a hold of $0.7$~s (black diamonds), without microwave field. The spectra are essentially identical; quantitatively, we extract $T/T_\F=0.24^{+7}_{-8}$ and $T/T_\F=0.23^{+7}_{-8}$ from $E_{\mathrm{p}} = -0.59(2) E_\F$ and $E_{\mathrm{p}} = -0.61(2) E_\F$ respectively. 

When the microwave induces rarefaction, the rf spectra significantly change (see magenta diamonds in Fig.~\ref{Fig:4}(c), corresponding to a rarefaction of $N/N_\mathrm{i}=0.28(6)$). In fact, we observe that the magnitude of $E_{\mathrm{p}}/E_{\mathrm{F}}$ continuously decreases with rarefaction (upper panel of Fig.~\ref{Fig:4}(d)), indicating qualitatively that the quantum degeneracy decreases. 
 In the lower panel of Fig.~\ref{Fig:4}(d) we show the evolution of $T/T_\F$ in a unitary-gas JT process. For an initial condition $\TTFi=0.24^{+7}_{-8}$ (blue diamonds), the data shows that the unitary gas experiences a weaker heating compared to the ideal Fermi gas (purple dash-dotted line); the data is in very good agreement with the prediction based on the experimentally measured EoS (blue solid line)~\cite{Ku_2012}.
 The band represents the uncertainty window arising from the uncertainty in $\TTFi$. 
We took an additional data set at a lower initial degeneracy, corresponding to $\TTFi=0.35(4)$, and observe weaker heating (see \cite{SuppMat} for details).

Despite the theoretical challenge in describing the strongly interacting Fermi gas, its JT effect is relatively simple to interpret in both the low-$T$ ($T\ll T_{\mathrm{c}}$) and high-$T$ ($T\gg T_\F$) limits.
In the low-$T$ limit, the unitary gas should exhibit a strong JT heating as $\theta_{\mathrm{JT}} \propto -\left(T/{T_\F}\right)^{-4}$ (which originates from both its non-vanishing ground state energy in the thermodynamic limit and its low-lying phononic excitations~\cite{ideallowT}). In 
 the high-$T$ limit, the unitary gas exhibits an effective interaction ($\propto -\log G(r)$) that is attractive~\cite{Ueff}; it should thus cool during a JT process~\cite{SuppMat}, akin to the ideal Bose gas. From the EoS, we expect that there exists an inversion temperature, \emph{i.e.} the temperature at which JT effect changes from heating to cooling, at $T_\mathrm{inv}\approx 0.9 T_{\F}$. In the intermediate range of $T/T_{\F}$ explored in this work, we observe weak heating, an effect in between the ideal Bose and Fermi gases (yellow and purple dot-dashed lines in Fig.\ref{Fig:4}(d)).

In conclusion, we realized JT processes in the essentially ideal Fermi gas and the unitary Fermi gas by exploiting scale invariance and implementing $u$-constant rarefactions. In the range of temperature explored, we observed JT heating in both cases and the effect is lessened when the repulsive quantum-statistical force is either weakened by the loss of degeneracy or counterbalanced by the attractive interparticle force. In the future, it would interesting to extend the study of the JT effect to the BEC-BCS crossover, where one expects a continuous transition from bosonic to fermionic behavior. While the absence of scale invariance poses an interesting experimental challenges on how to realize a JT process in that system, the JT coefficient could also be extracted from the isothermal compressibility~\cite{SuppMat,Ku_2012,mukherjee_2022}. The JT effect could also be observed in other interesting quantum many-body systems, such as dipolar gases~\cite{chomaz_2022}, low-dimensional and Hubbard systems~\cite{bloch_2012,gross_2021}.

We thank Fr\'{e}d\'{e}ric Chevy, Hadrien Kurkjian, Robert Smith, and  Zoran Hadzibabic for comments on the manuscript. We thank Martin Zwierlein and Biswaroop Mukherjee for sharing their data. This work was supported by the NSF (Grant Nos. PHY-1945324 and PHY-2110303), DARPA (Grant No. W911NF2010090), the David and Lucile Packard Foundation, and the Alfred P. Sloan Foundation. G.L.S. acknowledges  support  from  the  NSF  Graduate  Research  Fellowship  Program.

\newpage
\cleardoublepage
\setcounter{figure}{0}
\setcounter{equation}{0}
\newcounter{mycounter}
\setcounter{mycounter}{1}
\renewcommand{\thefigure}{S\arabic{figure}}
\newcommand{\RNum}[1]{\uppercase\expandafter{\romannumeral #1\relax.}}

\makeatletter
\def\maketitle{
\@author@finish
\title@column\titleblock@produce
\suppressfloats[t]}
\makeatother

\onecolumngrid

\title{Supplemental Material\\Observation of the Fermionic Joule-Thomson Effect}

\maketitle
\onecolumngrid

\section{\texorpdfstring{\RNum{\arabic{mycounter}}}. \texorpdfstring{$u$}{Lg}-constant versus \texorpdfstring{$h$}{Lg}-constant processes in a weakly interacting Fermi gas}
\stepcounter{mycounter}

Here we estimate the (small) difference between a $u$-constant rarefaction and a JT process in a weakly interacting Fermi gas, for which the relation $PV \propto U$ no longer holds. In the limit of weak interactions ($n|a|^3\ll 1$) and low temperatures  ($|a|/\lambda_{T} \ll 1$), the ratio $h/u$ is given to lowest order in $a$ by
\begin{equation}
\begin{split}
    \frac{h}{u} &= \frac{5\mathcal{F}_{5/2}(x_{\mathrm{id}}) + 4n^{\frac{1}{3}}a \left(\mathcal{F}_{3/2}(x_{\mathrm{id}})\right)^{\frac{5}{3}}}{3\mathcal{F}_{5/2}(x_{\mathrm{id}}) + 2n^{\frac{1}{3}}a \left(\mathcal{F}_{3/2}(x_{\mathrm{id}})\right)^{\frac{5}{3}}},
\end{split}\label{eq:weakEoS}
\end{equation}
where $\mathcal{F}_{j}(w) = \Gamma(j)^{-1} \int_0^{\infty} \dd y\;y^{j-1}/(\exp(y-w)+1)= -\mathrm{Li}_{j}(-\exp(w))$, $\mathrm{Li}_{j}$ is the polylogarithm of order $j$, $\Gamma$ is the gamma function, and $x_{\mathrm{id}}$ is the degeneracy parameter of a non-interacting Fermi gas at the same $n$ and $T$, \emph{i.e.} $\mathcal{F}_{3/2}(x_{\mathrm{id}}) = n \lambda_{T}^3$. 
In the range of interactions and temperatures explored in this work, we find that the specific enthalpy varies by  $|\Delta h|/h_{\mathrm{i}}\le 10^{-4}$ during the $u$-constant rarefaction ($h_{\mathrm{i}}$ is the initial specific enthalpy).

\section{\texorpdfstring{\RNum{\arabic{mycounter}}}. The Joule-Thomson coefficient}
\stepcounter{mycounter}
\subsection{Expressing \texorpdfstring{$\theta_\mathrm{JT}$}{Lg} in terms of the EoS}
Here we calculate the Joule-Thomson coefficient of a gas with a universal equation of state (EoS) of the form
\begin{equation}
        P(\mu,T) = \frac{k_{\B}T}{\lambda_{T}^3} f_P(x), 
\end{equation}
so that $n(\mu,T)\lambda_{T}^3 = f'_P(x)$,
where $x \equiv \mu/k_{\B} T$ ($\mu$ is the chemical potential). This form is valid for both ideal quantum gases and for the unitary Fermi gas. The temperature is $T/T_{\mathrm{F}} = (3 \sqrt{\pi}f'_P(x)/4)^{-2/3}$. For an ideal quantum gas, $f_P(x) = \eta \mathrm{Li}_{5/2}(\eta \exp(x))$ where $\eta = -1$ (resp. $\eta = 1$) for the ideal Fermi gas (resp. the ideal non-condensed Bose gas). Given $U = (3/2) PV$, the specific enthalpy is  
\begin{equation}
    \begin{split}
        h(\mu,T) &= \frac{5}{2} \frac{P}{n} = \frac{5}{2} k_{\B}T \frac{f_P(x)}{f'_P(x)}.
    \end{split} 
\end{equation}
Since $h$ is conserved in the JT process, the temperature evolution during the rarefaction satisfies the relation 

\begin{equation}
\begin{split}
        &x = \Lambda^{-1}\left(\frac{n}{n_{\mathrm{i}}} \Lambda(x_{\mathrm{i}})\right)  \\ 
        &\Lambda(x)=\frac{f'_P(x)^{\frac{5}{2}}}{f_P(x)^{\frac{3}{2}}},
\end{split} 
\end{equation}
where $x_{\mathrm{i}}=\mu_{\mathrm{i}}/k_\B T_{\mathrm{i}}$ ($\mu_{\mathrm{i}}$ and $T_{\mathrm{i}}$ are respectively the initial chemical potential and temperature). From this relation, we derive the solid lines in Fig.~2(d) and Fig.~4(d) in the main text.

\begin{figure}[h!]
\includegraphics[width=1\columnwidth]{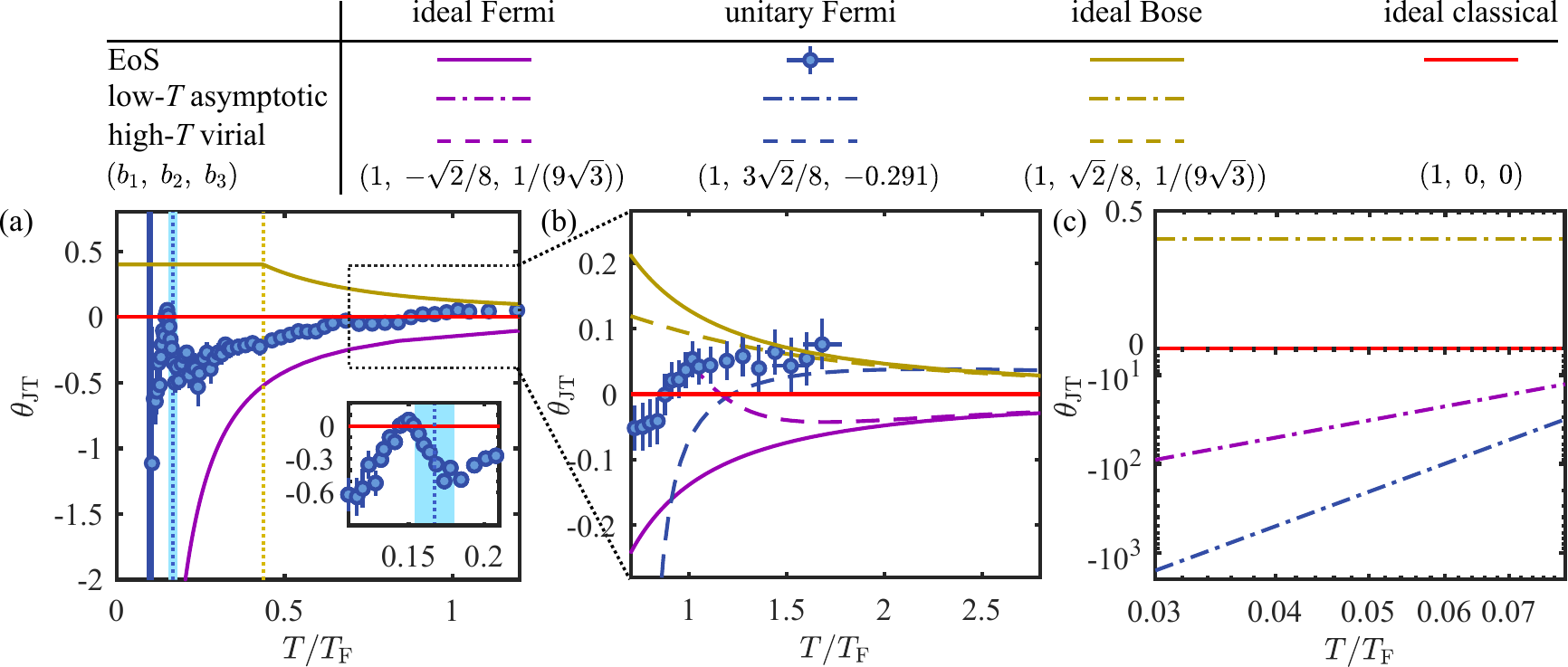}
\caption{Dimensionless Joule-Thomson coefficient $\theta_{\mathrm{JT}}$ in scale-invariant gases. (a) $\theta_{\mathrm{JT}}$ derived from EoS. The blue points are derived from the equation of state for the unitary Fermi gas~\cite{Ku_2012_2} (see also~\cite{mukherjee_2022_2}); the blue dotted line marks $T_{\mathrm{c}}$ for the unitary gas, and the band shows the uncertainty; the yellow dotted line marks $T_{\mathrm{c}}$ for the ideal Bose gas. (b) High-$T$ behavior of $\theta_{\mathrm{JT}}$. Dashed lines show the approximation using virial coefficients up to the third order~\cite{ho_2004,liu_2009}. (c) Low-$T$ behavior of $\theta_{\mathrm{JT}}$.}
\label{theta}
\end{figure}

Furthermore, we find $\mu_\mathrm{JT}$ to be
\begin{equation}
\label{theta_JT_i}
    \begin{split}
        \mu_{\mathrm{JT}} &\equiv \left(\frac{\partial T}{\partial P}\right)_h = -\left(\frac{\partial h}{\partial P}\right)_T \bigg/ \left(\frac{\partial h}{\partial T}\right)_P\\
        &= \frac{2}{5}\frac{\lambda_{T}^3}{k_{\B} f_P(x)} \frac{f_P(x)f''_P(x)-f'^2_P(x)}{f_P(x)f''_P(x) - \frac{3}{5}f'^2_P(x)}\\
    \end{split} 
\end{equation}
and $\theta_{\mathrm{JT}} =(P/T)\mu_{\mathrm{JT}}$. Fig.~\ref{theta}(a) shows $\theta_{\mathrm{JT}}$ for the ideal classical (red), Fermi (purple), and Bose gases (yellow), and the unitary Fermi gas (blue circles). 

The coefficient $\theta_{\mathrm{JT}}$ exhibits interesting features, namely first- and second-order discontinuities (for the unitary Fermi gas and ideal Bose gas, respectively, at their respective phase transition temperatures $T_{\mathrm{c}}$); below the critical temperature for Bose-Einstein condensation, it is constant,  $\theta_{\mathrm{JT}}=2/5$, as a result of the scaling between the critical density and temperature, $n_{\mathrm{c}} \propto T^{\frac{3}{2}}$. The unitary Fermi gas has an inversion temperature at high $T$ ($T \approx 0.9 T_{\F}$), as a result of the interplay between the quantum statistics and the attractive interparticle interactions.

\subsection{High-\texorpdfstring{$T$}{Lg} regime: virial coefficients and \texorpdfstring{$\theta_\mathrm{JT}$}{Lg}}

We discuss in the main text the link between the JT effect and interactions, either real ones or effective quantum-statistical ones. Here we quantify this relation in the high-$T$ limit using the virial expansion of the EoS, $P\lambda_T^3/(k_\B T) = \sum_{j = 1}^{\infty}  b_j(T) e^{jx}$, where $b_j$ is the $j$th-order virial coefficient. In Fig.~\ref{theta}(b), we show as dashed lines the high-$T$ approximation of $\theta_\mathrm{JT}$ up to $j=3$. If we instead truncate the expansion to $j=2$, the relation between $\theta_{\mathrm{JT}}$ and the coefficient $b_2$ can be simplified to
\begin{equation}
    \begin{split}
        \theta_{\mathrm{JT}} &\approx \frac{4}{3\sqrt{\pi}} \left(b_2(T)-\frac{2}{5} b_2'(T)T\right)\left(\frac{T}{T_{\F}}\right)^{-\frac{3}{2}},
    \end{split}
\end{equation}
where $b_2$ is related to the interaction potential $U_\mathrm{int}$ by 
\begin{equation}
    \begin{split}
        b_2(T) =\frac{2\pi}{\lambda_T^3} \int_{0}^{\infty} dr\;r^2 (e^{-U_\mathrm{int}(r)/(k_{\B}T)}-1).
    \end{split}
\end{equation}

For example, for a hard-sphere potential of radius $r_0$, one finds $ b_2(T) =-\frac{2\pi}{3\lambda_{T}^3}r_0^3$, so that $\theta_{\mathrm{JT}} < 0$.
We can get a more interesting $\theta_\mathrm{JT}$ by using a interaction potential model that has both short-range repulsion and long-range attraction. Specifically, let us pick $U_{\mathrm{int}}(r)=\infty$ for $r<r_0$ and $U_{\mathrm{int}}(r) = -C_6 \left(\frac{r_0}{r}\right)^6$ for $r\ge r_0$. The calculation can be carried out analytically:  
\begin{equation}
    \begin{split}
        \theta_{\mathrm{JT}} &\approx-\frac{8\sqrt{\pi}}{45\lambda_{T}^3}r_0^3 \left(2\exp\left(\frac{C_6}{k_{\B}T}\right) -3\sqrt{\frac{\pi C_6}{k_{\B}T}} \mathrm{Erfi} \left(\sqrt{\frac{C_6}{k_{\B}T}}\right)\right)\left(\frac{T}{T_{\F}}\right)^{-\frac{3}{2}}.
    \end{split} 
\end{equation}
where $\mathrm{Erfi}$ is the imaginary error function. We deduce from this expression that there is an inversion temperature $T_{\mathrm{inv}}\approx 2.3C_6/k_{\B}$, qualitatively capturing the essence of the JT effect in a classical interacting gas.

\begin{figure}[b]
\includegraphics[width=0.5\columnwidth]{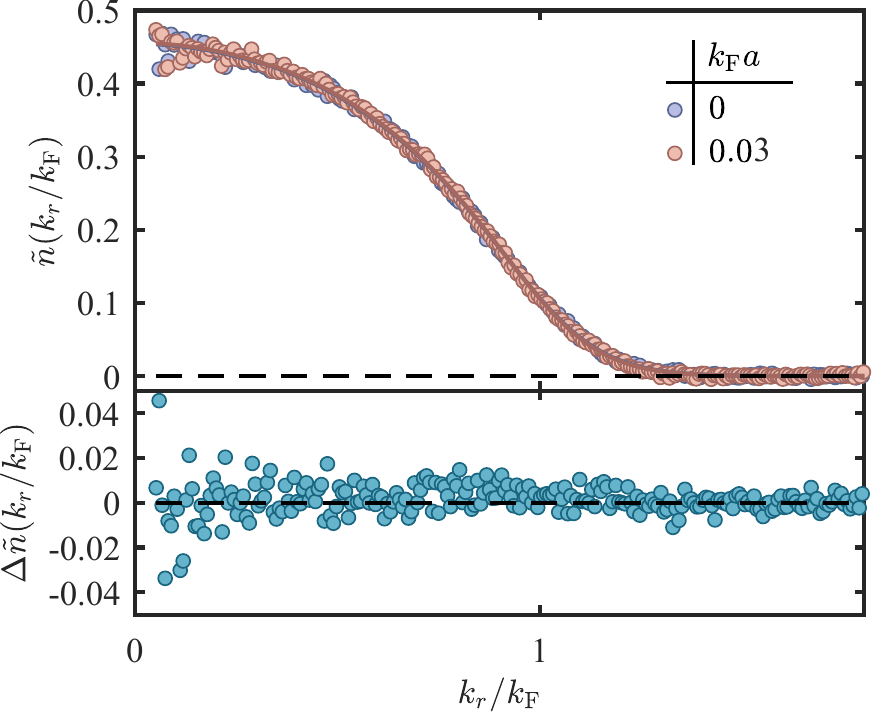}
\caption{Effect of interactions on the momentum distribution of a weakly interacting Fermi gas. The upper panel shows the momentum distribution of a weakly interacting Fermi gas measured before and after the interaction is suddenly turned off. The solid lines are fits to Fermi-Dirac distributions. The lower panel shows the difference between the two profiles.}
\label{momentum comparison}
\end{figure}

\subsection{Low-\texorpdfstring{$T$}{Lg} asymptote of \texorpdfstring{$\theta_\mathrm{JT}$}{Lg}}

In the low-$T$ limit, the energy per particle often takes the form
\begin{equation}
\begin{split}
    \frac{u}{\frac{3}{5}E_{\F}} &= A_g + A_e\Big(\frac{T}{T_\F}\Big)^{q},\\
\end{split}
\end{equation}
where the first term is the ground state energy and the second one is due to low-lying excitations. For the non-interacting Fermi gas, $q=2$ (corresponding to the particle-hole excitations); $q=4$ for the unitary Fermi gas (corresponding to its Bogoliubov-Anderson excitations); for the ideal Bose gas in the condensed phase, $q=5/2$. 
Assuming that $A_g$ and $A_e$ are not density-dependent and that $A_g \neq 0$, we find
\begin{equation}
 \theta_{\mathrm{JT}}\approx -\frac{2}{3q}\left(\frac{T}{T_\F}\right)^{-q}\left(\frac{A_g}{A_e}\right).
\end{equation}
If $A_g=0$ (e.g. in the ideal Bose gas case), $\theta_\text{JT}=(2/3)(1-1/q)$.
In Fig.~\ref{theta}(c), we show the asymptotes of $\theta_{\mathrm{JT}}$ for scale-invariant gases. Note that the JT heating effect in a unitary Fermi gas eventually becomes more pronounced than that of an ideal Fermi gas in the low-$T$ limit.

\section{\texorpdfstring{\RNum{\arabic{mycounter}}}. Momentum distribution of a weakly interacting gas}
\stepcounter{mycounter}
In our experiment, we perform time-of-flight thermometry on a weakly interacting gas ($k_{\F}a \le 0.02$) by fitting its momentum distribution to a Fermi-Dirac distribution. Here we show that the effect of interactions on the dynamics of the flight and the shape of the momentum distribution is negligible.

We prepare a weakly interacting gas at $583$~G ($a \approx 220 a_0$) and perform time-of-flight expansion at either the interaction field ($k_{\F}a\approx 0.03$), or the zero-crossing field $568$~G ($k_{\F}a \approx 0$). We compare the resulting momentum distributions in Fig.~\ref{momentum comparison}; the two measurements are essentially indistinguishable, with the same fitted temperature $T/T_{\F} = 0.22(1)$.

\section{\texorpdfstring{\RNum{\arabic{mycounter}}}. Box Trap Imperfections}
\stepcounter{mycounter}
\subsection{Technical heating}
\begin{figure}[h]
\includegraphics[width=1\columnwidth]{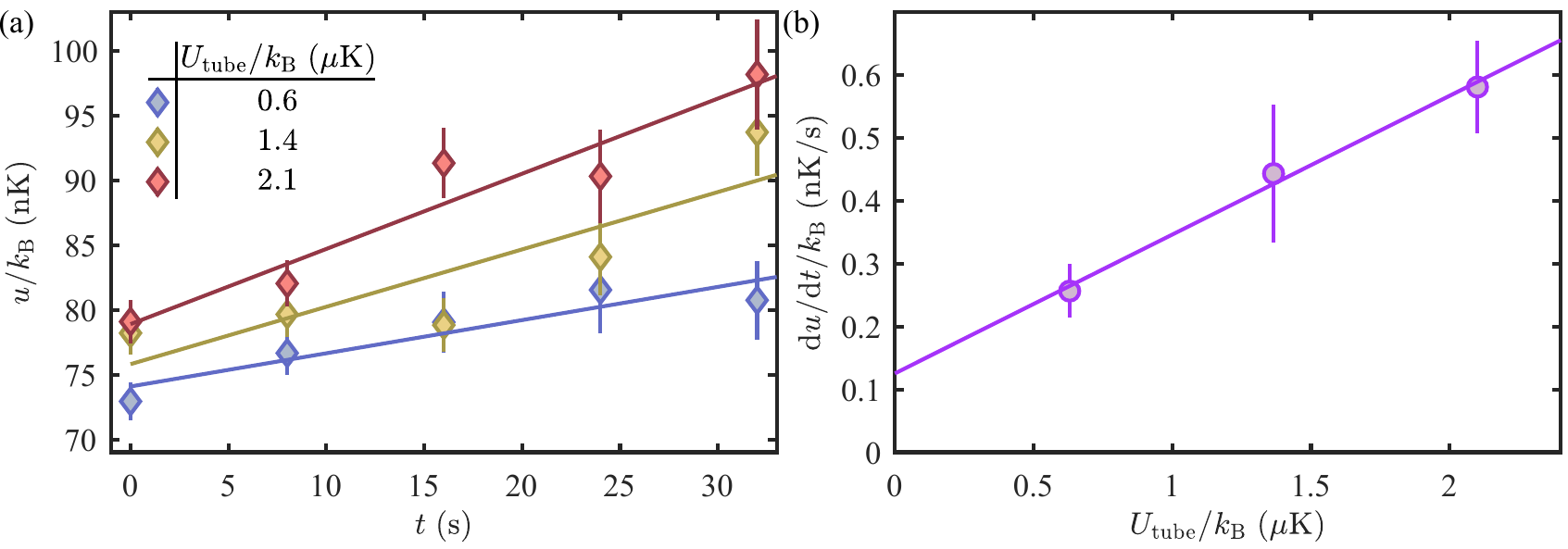}
\caption{Technical heating in the box trap. (a) Evolution of $u$ for different tube depths. Solid lines are linear fits to extract the heating rates. (b) Heating rate versus tube depth. The solid line is a linear fit.}
\label{technical_heating}
\end{figure}

We estimate the influence of technical heating on the measurement of the JT effect in our cylindrical box trap, which is constructed by intersecting a `tube' beam with two end `caps'. We measure the evolution of $u$ in a weakly interacting Fermi gas ($a \approx 220 a_0$). We ensure that the lifetime of the gas is vacuum limited at all depths of the tube (the tube being the dominant source of technical heating since at full power $U_{\mathrm{tube}} \ge 2U_{\mathrm{caps}}$) (Fig.~\ref{technical_heating}(a)). We find the technical heating rate to be $0.22(6) k_\B\times$~nK/s per $\mu\mathrm{K}$ of tube depth, and $\gamma_{\text{tech}}=0.58(7)k_\B\times$nK/s for the tube depth used in the experiments presented in the main text. 
The ideal gas $T$ dynamics including the effect of that heating shows improved agreement with the experimental data (dashed lines in Fig. 2(d) of the main text and Fig.~\ref{QJTIpow}).

\subsection{Imperfections of the trapping boundaries}

\begin{figure}[h]
\includegraphics[width=1\columnwidth]{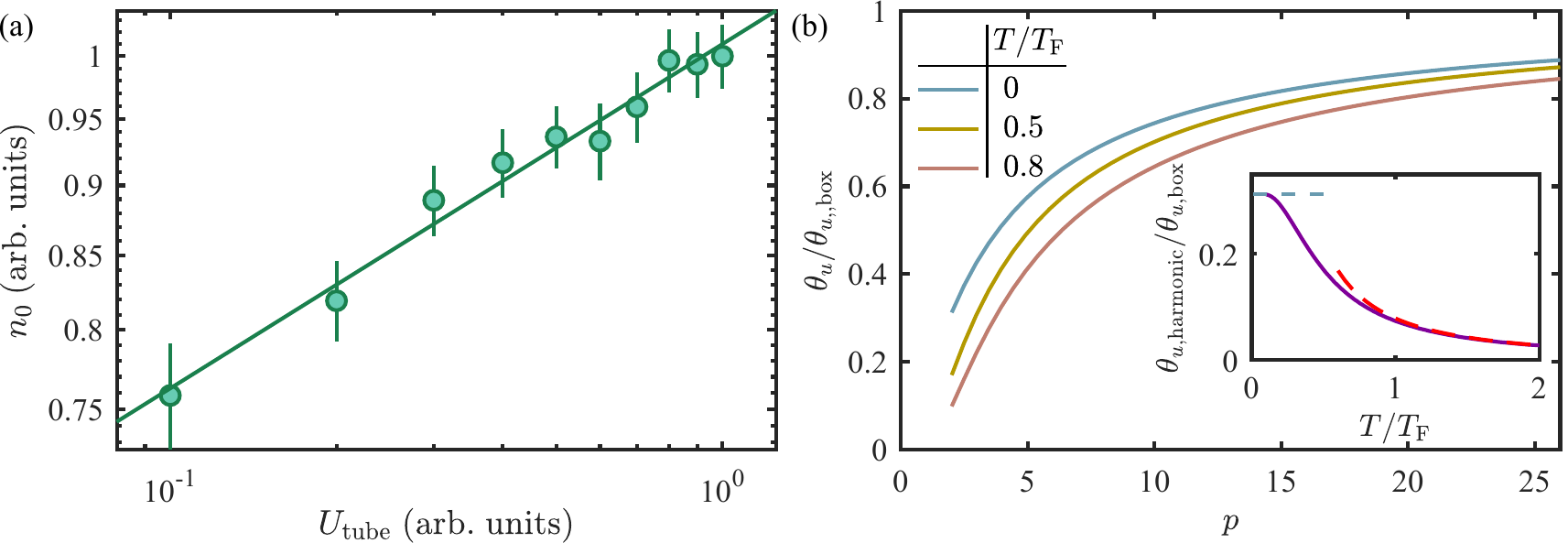}
\caption{Characterizing box trap imperfections and their effect on $\theta_u$. (a) Density of atoms at the center of the box trap $n_0$ as a function of the depth of the tube part of the box $U_\text{tube}$, for $N=9.9(7)\times 10^4$. The solid line is a power-law fit (see text). (b) $\theta_u$ of an ideal Fermi gas for various isotropic power-law trapping potentials, normalized to the perfect box trap value $\theta_{u,\text{box}}$ at the same $T/T_\F$. The blue line is the $T=0$ limit, $\theta_u/\theta_{u,\mathrm{box}}=15/(4 \nu(\nu+1))$. The inset shows $\theta_u$ in a harmonic trap. The blue dashed line shows the value at $T=0$, $\theta_{u,\mathrm{harmonic}}/\theta_{u,\mathrm{box}}=5/16$, and the red dashed line shows the high-$T$ limit approximation, $\theta_{u,\mathrm{harmonic}}/\theta_{u,\mathrm{box}}=(\sqrt{\pi/2}/16) (T/T_\F)^{-3/2}$.}
\label{boxgeo}
\end{figure}

Here we characterize the sharpness of our box trap walls, and assess its influence on the measurement of the JT effect in the weakly interacting Fermi gas. To address the non-uniform problem, we use $\theta_u \equiv (\partial \log(T)/\partial \log(N))_u$; in an ideal box-trapped Fermi gas $\theta_u = \theta_{\text{JT}}$. Because the tube part of our trap is (slightly) less sharp than the caps, we model the trapping potential as  $U_{\mathrm{b}}(\rho,z) = A\rho^p \Theta(z)\Theta(L-z)$ where $\rho$ and $z$ are the cylindrical coordinates with respect to the cylindrical box symmetry axis, and $\Theta$ is the Heaviside step function. Using the local density approximation, the density at the center of a fixed $N$, $T=0$ ideal Fermi gas is $n_0 \propto A^{6/(4+3p)}$. Experimentally, we load a weakly interacting gas in a shallow box, and control $A$ by raising the power of the tube beam (since $A\propto U_\text{tube}$). We access $n_0$ from \emph{in situ} imaging along the direction of tube. A power-law fit yields $p_{\mathrm{exp}}=15(4)$ (Fig.~\ref{boxgeo}(a)).

We estimate the effect of imperfect sharpness of the box on $\theta_u$. To make the calculation more tractable, we now assume an isotropic power-law potential $U_{\mathrm{b}} \propto r^{p}$, where $r$ is the spherical coordinate radius ($r=\sqrt{\rho^2+z^2}$). This model provides the worse-case scenario for the effect of the imperfect boundaries, given the exponent $p_\text{exp}$ determined above.

In this model, the energy per particle $u$, number of atoms $N$ and Fermi energy $E_\F$ of an ideal Fermi gas follow $u \propto T \mathcal{F}_{\nu+1}(x)/\mathcal{F}_{\nu}(x)$, $N \propto T^{\nu} \mathcal{F}_{\nu}(x)$, and $E_{\F}\propto N^{\frac{1}{\nu}}$, where $\nu =3/2 + 3/p$~\cite{li_1999,Faruk_2015}. We find  
\begin{equation}
    \begin{split}
        \theta_{u} &= \frac{\mathcal{F}_{\nu+1}(x)\mathcal{F}_{\nu-1}(x)-\mathcal{F}_{\nu}(x)^2}{(\nu+1)\mathcal{F}_{\nu+1}(x)\mathcal{F}_{\nu-1}(x)-\nu\mathcal{F}_{\nu}(x)^2}.
    \end{split} 
    \label{theta_JT_trap_general}
\end{equation}
 In the limit $p \rightarrow \infty$, Eq.~S\ref{theta_JT_trap_general} recovers Eq.~S\ref{theta_JT_i}.
 
\begin{figure}[h!]
\includegraphics[width=1\columnwidth]{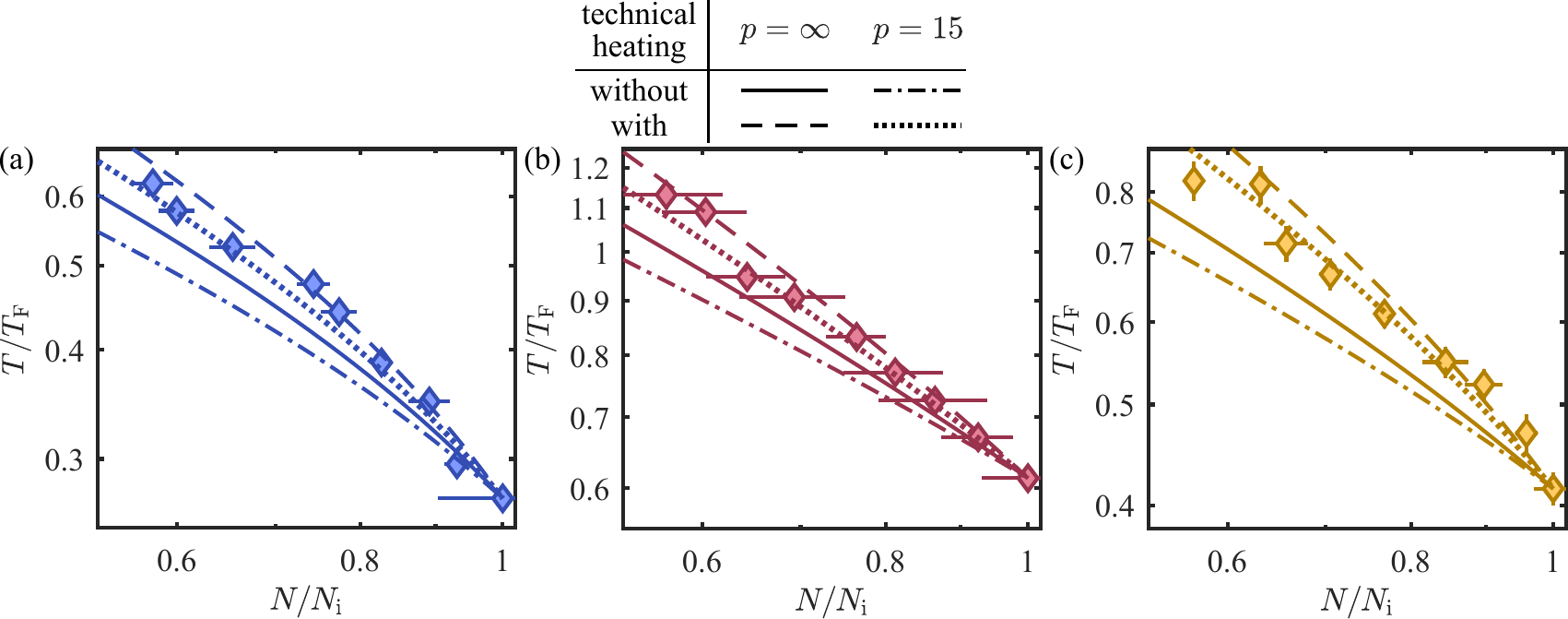}
\caption{Effect of box trap imperfections on the $T$ dynamics during rarefaction. In the first two panels, the data points, the solid lines and the dashed lines are the same as in Fig.~4(d) of the main text. The third panel shows an additional data set for an initial condition $\TTFi=0.42(2)$. The dash-dotted lines are the predictions in a power-law trap with $p=15$ (see text); the dotted lines also include technical heating in the box.}
\label{QJTIpow}
\end{figure}

 In Fig.~\ref{boxgeo}(b) we show $\theta_u(p)$. In Fig.~\ref{QJTIpow}, we show the predictions of the $T$ dynamics using $p=15$ (dotted lines and dash-dotted lines, respectively including technical heating or not). We find that taking into account the box trap imperfections and technical heating leads to excellent agreement with the experimental data.

\section{\texorpdfstring{\RNum{\arabic{mycounter}}}. Radio-Frequency Thermometry of a Unitary Fermi gas}
\stepcounter{mycounter}
\subsection{Linear response}

We verify that our spectroscopic thermometry is in the linear response regime. 
We perform the rf spectroscopy of a strongly interacting mixture of states $\ket{1}$-$\ket{3}$ by driving atoms on the transition $\ket{1}\rightarrow\ket{2}$ with a square pulse of Rabi frequency $\Omega_0$ and duration $t_\mathrm{pulse}$, and by measuring the resulting transferred fraction $N_2/N_1$. 
These rf spectra have been measured as a function of $T/T_{\F}$ at MIT~\cite{mukherjee_2019_2,yan_2022_2}, where the normalized response amplitude is independent of the rf parameters, \emph{i.e.} $N_{2}/N_{1} \propto t_{\mathrm{pulse}}\Omega_{0}^2$.

\begin{figure}[!hbt]
\includegraphics[width=1\columnwidth]{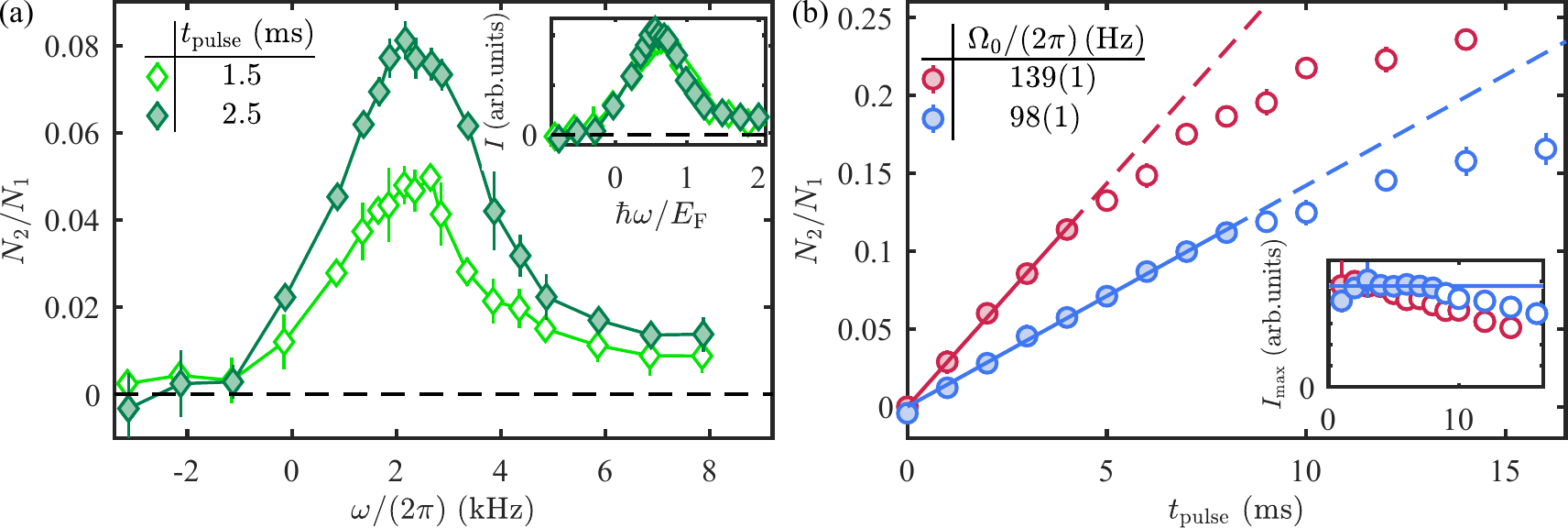}
\caption{Linear response of the radio-frequency  spectroscopy. (a) Transferred fraction $N_2/N_1$ as a function of the frequency detuning $\omega$ for two different pulse durations (and at the same power). Inset: Normalized spectral response. (b) Time-resolved transferred fraction $N_2/N_1$ at the peak response frequency $\omega_{\mathrm{p}}\approx 2\pi \times 2.3~\text{kHz}=0.58 E_\F/\hbar$. Solid-to-dashed lines are linear fits. Empty points are not included in the fitting. Inset: Normalized peak response amplitude. Solid lines correspond to the slopes of the linear fits in (b), and the bands show the uncertainties of the fits.}
\label{lr}
\end{figure}

We show in Fig.~\ref{lr}(a), two rf spectra of the same unitary gas ($E_\F/k_{\B} \approx 200$~nK) with two different pulse times ($t_{\mathrm{pulse}}=1.5$~ms and $t_{\mathrm{pulse}}=3$~ms) and the same $\Omega_0 = 2\pi\times 139(1)$~Hz. We find the normalized spectral responses to be nearly identical (inset of Fig.~\ref{lr}(a)). In Fig.~\ref{lr}(b), we show the time-resolved transferred fraction at the peak response frequency for the data shown in (a) ($\omega_{\mathrm{p}}\approx 2\pi \times 2.3~\text{kHz}=0.58 E_\F/\hbar$) for two different $\Omega_0$. In the inset of Fig.~\ref{lr}(b), we show that the normalized amplitude is the same for both $\Omega_0$ and independent of $t_\mathrm{pulse}$ for pulses in the short time limit.

\subsection{Peak frequency response thermometry}

\begin{figure}[h]
\includegraphics[width=1\columnwidth]{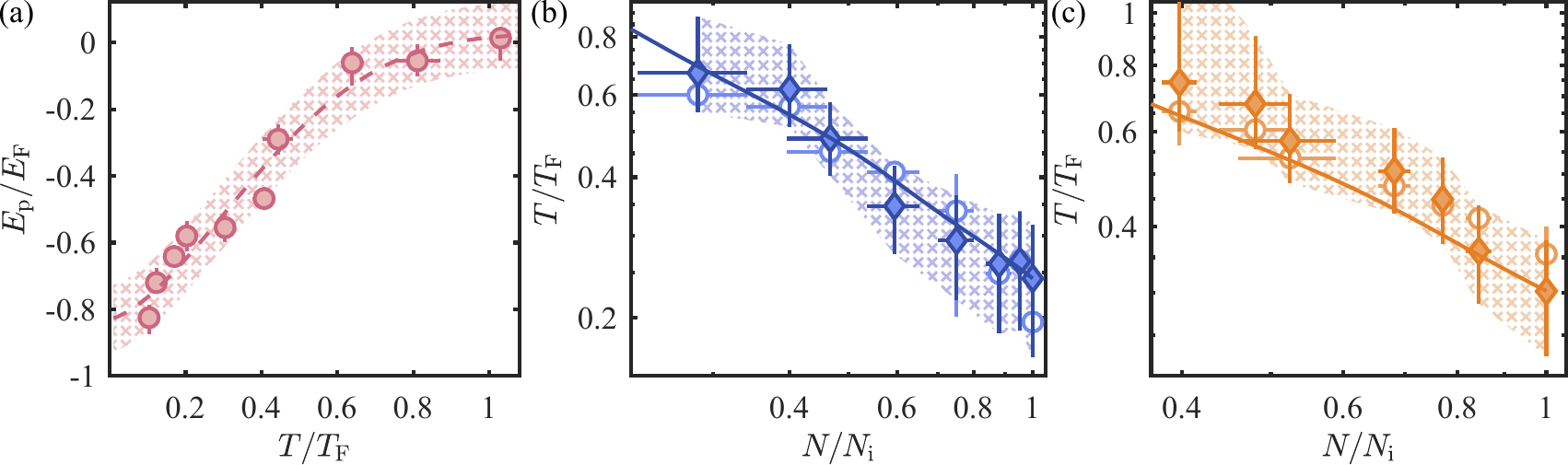}
\caption{Uncertainties on the rf thermometry. (a) $E_{\mathrm{p}}$ as a function of $T/T_\F$. The circles are the experimental measurements~\cite{mukherjee_2019_2}. The dashed line is a polynomial fit to the data and the band shows the estimated uncertainty. (b)-(c) $T$ dynamics in a JT process for two different $\TTFi$. The diamonds are extracted from the dashed line in (a). The error bars of the diamonds and bands correspond to the band in (a). Circles are directly interpolated from the experimental data. The solid lines are the theoretical predictions for the JT dynamics of the unitary Fermi gas.}
\label{Ep_robust}
\end{figure}

Here we estimate uncertainties on the rf thermometry using the peak frequency response $E_\mathrm{p}$. In Fig.~\ref{Ep_robust}(a) we show the experimental data of $E_\mathrm{p}$~\cite{mukherjee_2019_2} (circles).  We fit the data with a polynomial function (dashed line in Fig.~\ref{Ep_robust}(a)), and bound the data by shifting the fitted function (hatched band), from which we deduce the uncertainty bands for $T/T_\F$ (diamonds and hatched bands in Fig.~\ref{Ep_robust}(b-c)). The values of $T/T_\F$ extracted by direct interpolation of experimental data are shown as circles. 

\section{\texorpdfstring{\RNum{\arabic{mycounter}}}.  microwave transfer}
\stepcounter{mycounter}

\subsection{Energy-independent transfer}

\begin{figure}[!h]
\includegraphics[width=1\columnwidth]{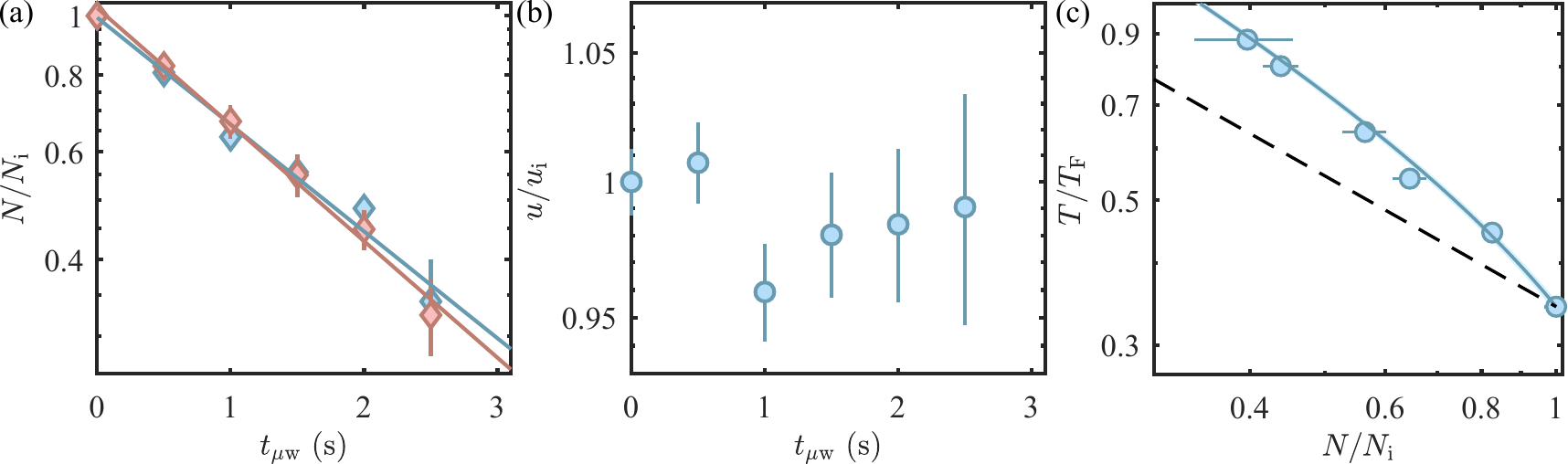}
\caption{Energy independence of the microwave-driven rarefaction of a weakly interacting Fermi gas. (a)  Decay of the gas as a function of the  microwave pulse duration $t_{\mu\mathrm{w}}$. Pink and blue diamonds are the populations in state $\ket{1}$ and $\ket{3}$ respectively. (b) Evolution of $u$ during  microwave transfer (normalized to the initial energy per particle $u_{\mathrm{i}}$). (c) Evolution of $T/T_\F$. The solid line is the theoretical prediction for the JT effect in an ideal Fermi gas; the band represents the uncertainty due to the error bar on $\TTFi$. The dashed line is the constant-$T$ reference.}
\label{uwave_check}
\end{figure}

 We verify that the  microwave-driven rarefaction (by transfer to the higher Zeeman sublevels) is energy independent by using the weakly interacting gas as a benchmark. We prepare a spin-balanced mixture of $\ket{1}$ and $\ket{3}$ at $582$~G ($a \approx 220 a_0$) and apply a two-tone  microwave pulse to induce equal losses in the two spin populations. In Fig.~\ref{uwave_check}(a) we show the decays of atom number, which correspond to lifetimes of $2.5(2)$~s and $2.3(2)$~s for states $\ket{1}$ and $\ket{3}$ respectively. After the  microwave transfer, we let the atoms thermalize for an additional $2$~s, during which no atom loss is observed (here $\Gamma_{\mathrm{el}} \ge 3.4\;\mathrm{s}^{-1}$). We subsequently extract $u$ from time-of-flight expansions and find that $u$ is constant during the  microwave rarefaction (Fig.~\ref{uwave_check}(b)). In addition, we show in Fig.~\ref{uwave_check}(c) the evolution of $T/T_{\F}$ during this process. Because the rarefaction time is now much shorter than the one induced by the collisions with the background particles, the technical heating is negligible on this time scale, and the data agrees very well with the theoretical prediction without taking the technical heating into account.

\subsection{Final state effects of the microwave transfer}

At $B\approx690$~G, the scattering lengths between the initial ($\ket{1}$ and $\ket{3}$) and final states ($\ket{4}$ and $\ket{6}$) used in the microwave transfer are small ($a_{\text{if}} \leq 50a_0$) for all combinations \cite{Lysebo_2009_2,kokkelmans}. To estimate the upper bound of the elastic scattering rate of the final states on the initial ones, $\Gamma_{\mathrm{if}}$, we assume that all transferred atoms are trapped and that their average velocity satisfies $\langle v_{\mathrm{f}} \rangle \le \sqrt{2U_{\mathrm{box}}/m}$ where $U_{\mathrm{box}}/k_{\B}\approx 1$~$\mu\mathrm{K}$ is the trap depth. We find $\Gamma_{\mathrm{if}} \le 0.8 \; \text{s}^{-1} \ll 1/\tau_{\mu\mathrm{w}}$. Therefore, the heat transfer between the initial and final states is negligible in our experiment.


\begin{thebibliography}{30}%
\makeatletter
\providecommand \@ifxundefined [1]{%
 \@ifx{#1\undefined}
}%
\providecommand \@ifnum [1]{%
 \ifnum #1\expandafter \@firstoftwo
 \else \expandafter \@secondoftwo
 \fi
}%
\providecommand \@ifx [1]{%
 \ifx #1\expandafter \@firstoftwo
 \else \expandafter \@secondoftwo
 \fi
}%
\providecommand \natexlab [1]{#1}%
\providecommand \enquote  [1]{``#1''}%
\providecommand \bibnamefont  [1]{#1}%
\providecommand \bibfnamefont [1]{#1}%
\providecommand \citenamefont [1]{#1}%
\providecommand \href@noop [0]{\@secondoftwo}%
\providecommand \href [0]{\begingroup \@sanitize@url \@href}%
\providecommand \@href[1]{\@@startlink{#1}\@@href}%
\providecommand \@@href[1]{\endgroup#1\@@endlink}%
\providecommand \@sanitize@url [0]{\catcode `\\12\catcode `\$12\catcode
  `\&12\catcode `\#12\catcode `\^12\catcode `\_12\catcode `\%12\relax}%
\providecommand \@@startlink[1]{}%
\providecommand \@@endlink[0]{}%
\providecommand \url  [0]{\begingroup\@sanitize@url \@url }%
\providecommand \@url [1]{\endgroup\@href {#1}{\urlprefix }}%
\providecommand \urlprefix  [0]{URL }%
\providecommand \Eprint [0]{\href }%
\providecommand \doibase [0]{https://doi.org/}%
\providecommand \selectlanguage [0]{\@gobble}%
\providecommand \bibinfo  [0]{\@secondoftwo}%
\providecommand \bibfield  [0]{\@secondoftwo}%
\providecommand \translation [1]{[#1]}%
\providecommand \BibitemOpen [0]{}%
\providecommand \bibitemStop [0]{}%
\providecommand \bibitemNoStop [0]{.\EOS\space}%
\providecommand \EOS [0]{\spacefactor3000\relax}%
\providecommand \BibitemShut  [1]{\csname bibitem#1\endcsname}%
\let\auto@bib@innerbib\@empty
%</preamble>
\bibitem [{\citenamefont {Joule}\ and\ \citenamefont
  {Thomson}(1852)}]{joule_1852}%
  \BibitemOpen
  \bibfield  {author} {\bibinfo {author} {\bibfnamefont {J.~P.}\ \bibnamefont
  {Joule}}\ and\ \bibinfo {author} {\bibfnamefont {W.}~\bibnamefont
  {Thomson}},\ }\href {https://doi.org/10.1080/14786445208647169} {\bibfield
  {journal} {\bibinfo  {journal} {Lond. Edinb. Dubl. Phil. Mag.}\ }\textbf
  {\bibinfo {volume} {4}},\ \bibinfo {pages} {481} (\bibinfo {year}
  {1852})}\BibitemShut {NoStop}%
\bibitem [{\citenamefont {Linde}(1899)}]{linde_1899}%
  \BibitemOpen
  \bibfield  {author} {\bibinfo {author} {\bibfnamefont {C.}~\bibnamefont
  {Linde}},\ }\href@noop {} {\bibfield  {journal} {\bibinfo  {journal}
  {Berichte der deutschen chemischen Gesellschaft}\ }\textbf {\bibinfo {volume}
  {32}},\ \bibinfo {pages} {925} (\bibinfo {year} {1899})}\BibitemShut
  {NoStop}%
\bibitem [{\citenamefont {Roebuck}(1926)}]{roebuck_1926}%
  \BibitemOpen
  \bibfield  {author} {\bibinfo {author} {\bibfnamefont {J.~R.}\ \bibnamefont
  {Roebuck}},\ }\href {https://doi.org/10.1073/pnas.12.1.55} {\bibfield
  {journal} {\bibinfo  {journal} {Proc. Natl. Acad. Sci. U.S.A.}\ }\textbf
  {\bibinfo {volume} {12}},\ \bibinfo {pages} {55} (\bibinfo {year}
  {1926})}\BibitemShut {NoStop}%
\bibitem [{\citenamefont {Hirschfelder}\ \emph {et~al.}(1938)\citenamefont
  {Hirschfelder}, \citenamefont {Ewell},\ and\ \citenamefont
  {Roebuck}}]{hirschfelder_1938}%
  \BibitemOpen
  \bibfield  {author} {\bibinfo {author} {\bibfnamefont {J.~O.}\ \bibnamefont
  {Hirschfelder}}, \bibinfo {author} {\bibfnamefont {R.~B.}\ \bibnamefont
  {Ewell}},\ and\ \bibinfo {author} {\bibfnamefont {J.~R.}\ \bibnamefont
  {Roebuck}},\ }\href {https://doi.org/doi.org/10.1063/1.1750228} {\bibfield
  {journal} {\bibinfo  {journal} {J. Chem. Phys.}\ }\textbf {\bibinfo {volume}
  {6}},\ \bibinfo {pages} {205} (\bibinfo {year} {1938})}\BibitemShut {NoStop}%
\bibitem [{\citenamefont {{\"O}kc{\"u}}\ and\ \citenamefont
  {Ayd{\i}ner}(2017)}]{Okcu_2017}%
  \BibitemOpen
  \bibfield  {author} {\bibinfo {author} {\bibfnamefont {{\"O}.}~\bibnamefont
  {{\"O}kc{\"u}}}\ and\ \bibinfo {author} {\bibfnamefont {E.}~\bibnamefont
  {Ayd{\i}ner}},\ }\href {https://doi.org/10.1140/epjc/s10052-017-4598-y}
  {\bibfield  {journal} {\bibinfo  {journal} {Eur. Phys. J. C}\ }\textbf
  {\bibinfo {volume} {77}},\ \bibinfo {pages} {1} (\bibinfo {year}
  {2017})}\BibitemShut {NoStop}%
\bibitem [{\citenamefont {Ghaffarnejad}\ \emph {et~al.}(2018)\citenamefont
  {Ghaffarnejad}, \citenamefont {Yaraie},\ and\ \citenamefont
  {Farsam}}]{ghaffarnejad_2018}%
  \BibitemOpen
  \bibfield  {author} {\bibinfo {author} {\bibfnamefont {H.}~\bibnamefont
  {Ghaffarnejad}}, \bibinfo {author} {\bibfnamefont {E.}~\bibnamefont
  {Yaraie}},\ and\ \bibinfo {author} {\bibfnamefont {M.}~\bibnamefont
  {Farsam}},\ }\href {https://doi.org/10.1007/s10773-018-3693-7} {\bibfield
  {journal} {\bibinfo  {journal} {Int. J. Theor. Phys.}\ }\textbf {\bibinfo
  {volume} {57}},\ \bibinfo {pages} {1671} (\bibinfo {year}
  {2018})}\BibitemShut {NoStop}%
\bibitem [{\citenamefont {Mo}\ \emph {et~al.}(2018)\citenamefont {Mo},
  \citenamefont {Li}, \citenamefont {Lan},\ and\ \citenamefont {Xu}}]{mo_2018}%
  \BibitemOpen
  \bibfield  {author} {\bibinfo {author} {\bibfnamefont {J.-X.}\ \bibnamefont
  {Mo}}, \bibinfo {author} {\bibfnamefont {G.-Q.}\ \bibnamefont {Li}}, \bibinfo
  {author} {\bibfnamefont {S.-Q.}\ \bibnamefont {Lan}},\ and\ \bibinfo {author}
  {\bibfnamefont {X.-B.}\ \bibnamefont {Xu}},\ }\href
  {https://doi.org/10.1103/PhysRevD.98.124032} {\bibfield  {journal} {\bibinfo
  {journal} {Phys. Rev. D}\ }\textbf {\bibinfo {volume} {98}},\ \bibinfo
  {pages} {124032} (\bibinfo {year} {2018})}\BibitemShut {NoStop}%
\bibitem [{\citenamefont {Kothari}\ and\ \citenamefont
  {Srivasava}(1937)}]{kothari_1937}%
  \BibitemOpen
  \bibfield  {author} {\bibinfo {author} {\bibfnamefont {D.~S.}\ \bibnamefont
  {Kothari}}\ and\ \bibinfo {author} {\bibfnamefont {B.~N.}\ \bibnamefont
  {Srivasava}},\ }\href {https://doi.org/pdf/10.1098/rspa.1939.0079} {\bibfield
   {journal} {\bibinfo  {journal} {Nature}\ }\textbf {\bibinfo {volume}
  {140}},\ \bibinfo {pages} {970} (\bibinfo {year} {1937})}\BibitemShut
  {NoStop}%
\bibitem [{\citenamefont {Schmidutz}\ \emph {et~al.}(2014)\citenamefont
  {Schmidutz}, \citenamefont {Gotlibovych}, \citenamefont {Gaunt},
  \citenamefont {Smith}, \citenamefont {Navon},\ and\ \citenamefont
  {Hadzibabic}}]{Schmidutz_2014}%
  \BibitemOpen
  \bibfield  {author} {\bibinfo {author} {\bibfnamefont {T.~F.}\ \bibnamefont
  {Schmidutz}}, \bibinfo {author} {\bibfnamefont {I.}~\bibnamefont
  {Gotlibovych}}, \bibinfo {author} {\bibfnamefont {A.~L.}\ \bibnamefont
  {Gaunt}}, \bibinfo {author} {\bibfnamefont {R.~P.}\ \bibnamefont {Smith}},
  \bibinfo {author} {\bibfnamefont {N.}~\bibnamefont {Navon}},\ and\ \bibinfo
  {author} {\bibfnamefont {Z.}~\bibnamefont {Hadzibabic}},\ }\href
  {https://doi.org/10.1103/PhysRevLett.112.040403} {\bibfield  {journal}
  {\bibinfo  {journal} {Phys. Rev. Lett.}\ }\textbf {\bibinfo {volume} {112}},\
  \bibinfo {pages} {040403} (\bibinfo {year} {2014})}\BibitemShut {NoStop}%
\bibitem [{Sup()}]{SuppMat}%
  \BibitemOpen
  \href@noop {} {}\bibinfo {note} {See Supplementary Material.}\BibitemShut
  {Stop}%
\bibitem [{\citenamefont {Navon}\ \emph {et~al.}(2021)\citenamefont {Navon},
  \citenamefont {Smith},\ and\ \citenamefont {Hadzibabic}}]{navon_2021}%
  \BibitemOpen
  \bibfield  {author} {\bibinfo {author} {\bibfnamefont {N.}~\bibnamefont
  {Navon}}, \bibinfo {author} {\bibfnamefont {R.~P.}\ \bibnamefont {Smith}},\
  and\ \bibinfo {author} {\bibfnamefont {Z.}~\bibnamefont {Hadzibabic}},\
  }\href {https://doi.org/10.1038/s41567-021-01403-z} {\bibfield  {journal}
  {\bibinfo  {journal} {Nat. Phys.}\ }\textbf {\bibinfo {volume} {17}},\
  \bibinfo {pages} {1334} (\bibinfo {year} {2021})}\BibitemShut {NoStop}%
\bibitem [{\citenamefont {Timmermans}(2001)}]{Timmermans_2001}%
  \BibitemOpen
  \bibfield  {author} {\bibinfo {author} {\bibfnamefont {E.}~\bibnamefont
  {Timmermans}},\ }\href {https://doi.org/10.1103/PhysRevLett.87.240403}
  {\bibfield  {journal} {\bibinfo  {journal} {Phys. Rev. Lett.}\ }\textbf
  {\bibinfo {volume} {87}},\ \bibinfo {pages} {240403} (\bibinfo {year}
  {2001})}\BibitemShut {NoStop}%
\bibitem [{\citenamefont {Pathria}\ and\ \citenamefont
  {Beale}(2011)}]{pathria_2011}%
  \BibitemOpen
  \bibfield  {author} {\bibinfo {author} {\bibfnamefont {R.~K.}\ \bibnamefont
  {Pathria}}\ and\ \bibinfo {author} {\bibfnamefont {P.~D.}\ \bibnamefont
  {Beale}},\ }\href@noop {} {\emph {\bibinfo {title} {Statistical Mechanics 3rd
  ed.}}}\ (\bibinfo  {publisher} {Elsevier},\ \bibinfo {year}
  {2011})\BibitemShut {NoStop}%
\bibitem [{\citenamefont {Uhlenbeck}\ and\ \citenamefont
  {Gropper}(1932)}]{uhlenbeck_1932}%
  \BibitemOpen
  \bibfield  {author} {\bibinfo {author} {\bibfnamefont {G.}~\bibnamefont
  {Uhlenbeck}}\ and\ \bibinfo {author} {\bibfnamefont {L.}~\bibnamefont
  {Gropper}},\ }\href {https://doi.org/10.1103/PhysRev.41.79} {\bibfield
  {journal} {\bibinfo  {journal} {Phys. Rev.}\ }\textbf {\bibinfo {volume}
  {41}},\ \bibinfo {pages} {79} (\bibinfo {year} {1932})}\BibitemShut {NoStop}%
\bibitem [{\citenamefont {Mullin}\ and\ \citenamefont
  {Blaylock}(2003)}]{mullin_2003}%
  \BibitemOpen
  \bibfield  {author} {\bibinfo {author} {\bibfnamefont {W.~J.}\ \bibnamefont
  {Mullin}}\ and\ \bibinfo {author} {\bibfnamefont {G.}~\bibnamefont
  {Blaylock}},\ }\href {https://doi.org/10.1119/1.1590658} {\bibfield
  {journal} {\bibinfo  {journal} {Am. J. Phys.}\ }\textbf {\bibinfo {volume}
  {71}},\ \bibinfo {pages} {1223} (\bibinfo {year} {2003})}\BibitemShut
  {NoStop}%
\bibitem [{vir()}]{virial}%
  \BibitemOpen
  \href@noop {} {}\bibinfo {note} {The relation between the JT effect and
  two-body interactions can be quantitatively specified at the level of the
  second-order virial coefficient. Indeed, $\theta_{\text{JT}} \propto
  b_2(T)-\frac{2}{5} b_2'(T)T$, where $b_2 = \int \mathrm{d} \mathbf{r_1}
  \mathrm{d} \mathbf{r_2}
  (1-G(\mathbf{r_1},\mathbf{r_2}))/(2V\eta\lambda_T^3)$~\cite{SuppMat}.}\BibitemShut
  {Stop}%
\bibitem [{\citenamefont {Tan}(2008)}]{Tan_2008_Virial}%
  \BibitemOpen
  \bibfield  {author} {\bibinfo {author} {\bibfnamefont {S.}~\bibnamefont
  {Tan}},\ }\href {https://doi.org/10.1016/j.aop.2008.03.003} {\bibfield
  {journal} {\bibinfo  {journal} {Ann. Phys. (N. Y.)}\ }\textbf {\bibinfo
  {volume} {323}},\ \bibinfo {pages} {2987} (\bibinfo {year}
  {2008})}\BibitemShut {NoStop}%
\bibitem [{\citenamefont {Zwerger}(2011)}]{zwerger_2011}%
  \BibitemOpen
  \bibfield  {author} {\bibinfo {author} {\bibfnamefont {W.}~\bibnamefont
  {Zwerger}},\ }\href@noop {} {\emph {\bibinfo {title} {The BCS-BEC crossover
  and the unitary Fermi gas}}},\ Vol.\ \bibinfo {volume} {836}\ (\bibinfo
  {publisher} {Springer Science \& Business Media},\ \bibinfo {year}
  {2011})\BibitemShut {NoStop}%
\bibitem [{uni()}]{unitarylifetime}%
  \BibitemOpen
  \href@noop {} {}\bibinfo {note} {We observe that $\tau_\text{uni}$ saturates
  to a value shorter than $\tau_1$ with increasing $U_{\text{box}}$, indicating
  that a loss mechanism other than evaporation is taking place}\BibitemShut
  {NoStop}%
\bibitem [{mic()}]{microwave}%
  \BibitemOpen
  \href@noop {} {}\bibinfo {note} {The absorption of the microwave photons is
  independent of the atoms' energy because the Doppler effect is negligible. In
  addition, the interactions of atoms in the final states with those in the
  initial states are very weak~\cite{Lysebo_2009}.}\BibitemShut {Stop}%
\bibitem [{\citenamefont {Ku}\ \emph {et~al.}(2012)\citenamefont {Ku},
  \citenamefont {Sommer}, \citenamefont {Cheuk},\ and\ \citenamefont
  {Zwierlein}}]{Ku_2012}%
  \BibitemOpen
  \bibfield  {author} {\bibinfo {author} {\bibfnamefont {M.~J.~H.}\
  \bibnamefont {Ku}}, \bibinfo {author} {\bibfnamefont {A.~T.}\ \bibnamefont
  {Sommer}}, \bibinfo {author} {\bibfnamefont {L.~W.}\ \bibnamefont {Cheuk}},\
  and\ \bibinfo {author} {\bibfnamefont {M.~W.}\ \bibnamefont {Zwierlein}},\
  }\href {https://doi.org/10.1126/science.1214987} {\bibfield  {journal}
  {\bibinfo  {journal} {Science}\ }\textbf {\bibinfo {volume} {335}},\ \bibinfo
  {pages} {563} (\bibinfo {year} {2012})}\BibitemShut {NoStop}%
\bibitem [{\citenamefont {Mukherjee}\ \emph {et~al.}(2019)\citenamefont
  {Mukherjee}, \citenamefont {Patel}, \citenamefont {Yan}, \citenamefont
  {Fletcher}, \citenamefont {Struck},\ and\ \citenamefont
  {Zwierlein}}]{mukherjee_2019}%
  \BibitemOpen
  \bibfield  {author} {\bibinfo {author} {\bibfnamefont {B.}~\bibnamefont
  {Mukherjee}}, \bibinfo {author} {\bibfnamefont {P.~B.}\ \bibnamefont
  {Patel}}, \bibinfo {author} {\bibfnamefont {Z.}~\bibnamefont {Yan}}, \bibinfo
  {author} {\bibfnamefont {R.~J.}\ \bibnamefont {Fletcher}}, \bibinfo {author}
  {\bibfnamefont {J.}~\bibnamefont {Struck}},\ and\ \bibinfo {author}
  {\bibfnamefont {M.~W.}\ \bibnamefont {Zwierlein}},\ }\href
  {https://doi.org/10.1103/PhysRevLett.122.203402} {\bibfield  {journal}
  {\bibinfo  {journal} {Phys. Rev. Lett.}\ }\textbf {\bibinfo {volume} {122}},\
  \bibinfo {pages} {203402} (\bibinfo {year} {2019})}\BibitemShut {NoStop}%
\bibitem [{\citenamefont {Yan}\ \emph {et~al.}(2022)\citenamefont {Yan},
  \citenamefont {Patel}, \citenamefont {Mukherjee}, \citenamefont {Vale},
  \citenamefont {Fletcher},\ and\ \citenamefont {Zwierlein}}]{yan_2022}%
  \BibitemOpen
  \bibfield  {author} {\bibinfo {author} {\bibfnamefont {Z.}~\bibnamefont
  {Yan}}, \bibinfo {author} {\bibfnamefont {P.~B.}\ \bibnamefont {Patel}},
  \bibinfo {author} {\bibfnamefont {B.}~\bibnamefont {Mukherjee}}, \bibinfo
  {author} {\bibfnamefont {C.~J.}\ \bibnamefont {Vale}}, \bibinfo {author}
  {\bibfnamefont {R.~J.}\ \bibnamefont {Fletcher}},\ and\ \bibinfo {author}
  {\bibfnamefont {M.}~\bibnamefont {Zwierlein}},\ }\href@noop {} {} (\bibinfo
  {year} {2022}),\ \Eprint {https://arxiv.org/abs/2212.13752}
  {arXiv:2212.13752} \BibitemShut {NoStop}%
\bibitem [{ide()}]{ideallowT}%
  \BibitemOpen
  \href@noop {} {}\bibinfo {note} {For the ideal Fermi gas,
  $\theta_{\mathrm{JT}} \approx -\left(T/{T_\F}\right)^{-2}$ in the low-$T$
  limit; the quadratic dependence stems from the particle-hole nature of the
  excitations.}\BibitemShut {Stop}%
\bibitem [{Uef()}]{Ueff}%
  \BibitemOpen
  \href@noop {} {}\bibinfo {note} {Unlike in ideal quantum gases, the origin of
  the effective high-$T$ interaction potential of the unitary Fermi gas is not
  purely quantum statistical; it results from both the repulsive
  quantum-statistical force and the direct (effectively attractive) interatomic
  contact interaction.}\BibitemShut {Stop}%
\bibitem [{\citenamefont {Mukherjee}(2022)}]{mukherjee_2022}%
  \BibitemOpen
  \bibfield  {author} {\bibinfo {author} {\bibfnamefont {B.}~\bibnamefont
  {Mukherjee}},\ }\emph {\bibinfo {title} {Homogeneous quantum gases: strongly
  interacting fermions and rotating bosonic condensates}},\ \href@noop {}
  {Ph.D. thesis},\ \bibinfo  {school} {Massachusetts Institute of Technology}
  (\bibinfo {year} {2022})\BibitemShut {NoStop}%
\bibitem [{\citenamefont {Chomaz}\ \emph {et~al.}(2022)\citenamefont {Chomaz},
  \citenamefont {Ferrier-Barbut}, \citenamefont {Ferlaino}, \citenamefont
  {Laburthe-Tolra}, \citenamefont {Lev},\ and\ \citenamefont
  {Pfau}}]{chomaz_2022}%
  \BibitemOpen
  \bibfield  {author} {\bibinfo {author} {\bibfnamefont {L.}~\bibnamefont
  {Chomaz}}, \bibinfo {author} {\bibfnamefont {I.}~\bibnamefont
  {Ferrier-Barbut}}, \bibinfo {author} {\bibfnamefont {F.}~\bibnamefont
  {Ferlaino}}, \bibinfo {author} {\bibfnamefont {B.}~\bibnamefont
  {Laburthe-Tolra}}, \bibinfo {author} {\bibfnamefont {B.~L.}\ \bibnamefont
  {Lev}},\ and\ \bibinfo {author} {\bibfnamefont {T.}~\bibnamefont {Pfau}},\
  }\href {https://doi.org/10.1088/1361-6633/aca814} {\bibfield  {journal}
  {\bibinfo  {journal} {Rep. Prog. Phys.}\ }\textbf {\bibinfo {volume} {86}},\
  \bibinfo {pages} {026401} (\bibinfo {year} {2022})}\BibitemShut {NoStop}%
\bibitem [{\citenamefont {Bloch}\ \emph {et~al.}(2012)\citenamefont {Bloch},
  \citenamefont {Dalibard},\ and\ \citenamefont {Nascimbene}}]{bloch_2012}%
  \BibitemOpen
  \bibfield  {author} {\bibinfo {author} {\bibfnamefont {I.}~\bibnamefont
  {Bloch}}, \bibinfo {author} {\bibfnamefont {J.}~\bibnamefont {Dalibard}},\
  and\ \bibinfo {author} {\bibfnamefont {S.}~\bibnamefont {Nascimbene}},\
  }\href {https://doi.org/10.1038/nphys2259} {\bibfield  {journal} {\bibinfo
  {journal} {Nat. Phys.}\ }\textbf {\bibinfo {volume} {8}},\ \bibinfo {pages}
  {267} (\bibinfo {year} {2012})}\BibitemShut {NoStop}%
\bibitem [{\citenamefont {Gross}\ and\ \citenamefont
  {Bakr}(2021)}]{gross_2021}%
  \BibitemOpen
  \bibfield  {author} {\bibinfo {author} {\bibfnamefont {C.}~\bibnamefont
  {Gross}}\ and\ \bibinfo {author} {\bibfnamefont {W.~S.}\ \bibnamefont
  {Bakr}},\ }\href {https://doi.org/10.1038/s41567-021-01370-5} {\bibfield
  {journal} {\bibinfo  {journal} {Nat. Phys.}\ }\textbf {\bibinfo {volume}
  {17}},\ \bibinfo {pages} {1316} (\bibinfo {year} {2021})}\BibitemShut
  {NoStop}%
\bibitem [{\citenamefont {Lysebo}\ and\ \citenamefont
  {Veseth}(2009)}]{Lysebo_2009}%
  \BibitemOpen
  \bibfield  {author} {\bibinfo {author} {\bibfnamefont {M.}~\bibnamefont
  {Lysebo}}\ and\ \bibinfo {author} {\bibfnamefont {L.}~\bibnamefont
  {Veseth}},\ }\href {https://doi.org/10.1103/PhysRevA.79.062704} {\bibfield
  {journal} {\bibinfo  {journal} {Phys. Rev. A}\ }\textbf {\bibinfo {volume}
  {79}},\ \bibinfo {pages} {062704} (\bibinfo {year} {2009})}\BibitemShut
  {NoStop}%
\end{thebibliography}

\begin{thebibliography}{10}%
\makeatletter
\providecommand \@ifxundefined [1]{%
 \@ifx{#1\undefined}
}%
\providecommand \@ifnum [1]{%
 \ifnum #1\expandafter \@firstoftwo
 \else \expandafter \@secondoftwo
 \fi
}%
\providecommand \@ifx [1]{%
 \ifx #1\expandafter \@firstoftwo
 \else \expandafter \@secondoftwo
 \fi
}%
\providecommand \natexlab [1]{#1}%
\providecommand \enquote  [1]{``#1''}%
\providecommand \bibnamefont  [1]{#1}%
\providecommand \bibfnamefont [1]{#1}%
\providecommand \citenamefont [1]{#1}%
\providecommand \href@noop [0]{\@secondoftwo}%
\providecommand \href [0]{\begingroup \@sanitize@url \@href}%
\providecommand \@href[1]{\@@startlink{#1}\@@href}%
\providecommand \@@href[1]{\endgroup#1\@@endlink}%
\providecommand \@sanitize@url [0]{\catcode `\\12\catcode `\$12\catcode
  `\&12\catcode `\#12\catcode `\^12\catcode `\_12\catcode `\%12\relax}%
\providecommand \@@startlink[1]{}%
\providecommand \@@endlink[0]{}%
\providecommand \url  [0]{\begingroup\@sanitize@url \@url }%
\providecommand \@url [1]{\endgroup\@href {#1}{\urlprefix }}%
\providecommand \urlprefix  [0]{URL }%
\providecommand \Eprint [0]{\href }%
\providecommand \doibase [0]{https://doi.org/}%
\providecommand \selectlanguage [0]{\@gobble}%
\providecommand \bibinfo  [0]{\@secondoftwo}%
\providecommand \bibfield  [0]{\@secondoftwo}%
\providecommand \translation [1]{[#1]}%
\providecommand \BibitemOpen [0]{}%
\providecommand \bibitemStop [0]{}%
\providecommand \bibitemNoStop [0]{.\EOS\space}%
\providecommand \EOS [0]{\spacefactor3000\relax}%
\providecommand \BibitemShut  [1]{\csname bibitem#1\endcsname}%
\let\auto@bib@innerbib\@empty
%</preamble>
\bibitem [{\citenamefont {Ku}\ \emph {et~al.}(2012)\citenamefont {Ku},
  \citenamefont {Sommer}, \citenamefont {Cheuk},\ and\ \citenamefont
  {Zwierlein}}]{Ku_2012_2}%
  \BibitemOpen
  \bibfield  {author} {\bibinfo {author} {\bibfnamefont {M.~J.~H.}\
  \bibnamefont {Ku}}, \bibinfo {author} {\bibfnamefont {A.~T.}\ \bibnamefont
  {Sommer}}, \bibinfo {author} {\bibfnamefont {L.~W.}\ \bibnamefont {Cheuk}},\
  and\ \bibinfo {author} {\bibfnamefont {M.~W.}\ \bibnamefont {Zwierlein}},\
  }\href {https://doi.org/10.1126/science.1214987} {\bibfield  {journal}
  {\bibinfo  {journal} {Science}\ }\textbf {\bibinfo {volume} {335}},\ \bibinfo
  {pages} {563} (\bibinfo {year} {2012})}\BibitemShut {NoStop}%
\bibitem [{\citenamefont {Mukherjee}(2022)}]{mukherjee_2022_2}%
  \BibitemOpen
  \bibfield  {author} {\bibinfo {author} {\bibfnamefont {B.}~\bibnamefont
  {Mukherjee}},\ }\emph {\bibinfo {title} {Homogeneous quantum gases: strongly
  interacting fermions and rotating bosonic condensates}},\ \href@noop {}
  {Ph.D. thesis},\ \bibinfo  {school} {Massachusetts Institute of Technology}
  (\bibinfo {year} {2022})\BibitemShut {NoStop}%
\bibitem [{\citenamefont {Ho}\ and\ \citenamefont {Mueller}(2004)}]{ho_2004}%
  \BibitemOpen
  \bibfield  {author} {\bibinfo {author} {\bibfnamefont {T.-L.}\ \bibnamefont
  {Ho}}\ and\ \bibinfo {author} {\bibfnamefont {E.~J.}\ \bibnamefont
  {Mueller}},\ }\href {https://doi.org/10.1103/PhysRevLett.92.160404}
  {\bibfield  {journal} {\bibinfo  {journal} {Phys. Rev. Lett.}\ }\textbf
  {\bibinfo {volume} {92}},\ \bibinfo {pages} {160404} (\bibinfo {year}
  {2004})}\BibitemShut {NoStop}%
\bibitem [{\citenamefont {Liu}\ \emph {et~al.}(2009)\citenamefont {Liu},
  \citenamefont {Hu},\ and\ \citenamefont {Drummond}}]{liu_2009}%
  \BibitemOpen
  \bibfield  {author} {\bibinfo {author} {\bibfnamefont {X.-J.}\ \bibnamefont
  {Liu}}, \bibinfo {author} {\bibfnamefont {H.}~\bibnamefont {Hu}},\ and\
  \bibinfo {author} {\bibfnamefont {P.~D.}\ \bibnamefont {Drummond}},\ }\href
  {https://doi.org/10.1103/PhysRevLett.102.160401} {\bibfield  {journal}
  {\bibinfo  {journal} {Phys. Rev. Lett.}\ }\textbf {\bibinfo {volume} {102}},\
  \bibinfo {pages} {160401} (\bibinfo {year} {2009})}\BibitemShut {NoStop}%
\bibitem [{\citenamefont {Li}\ \emph {et~al.}(1999)\citenamefont {Li},
  \citenamefont {Chen},\ and\ \citenamefont {Chen}}]{li_1999}%
  \BibitemOpen
  \bibfield  {author} {\bibinfo {author} {\bibfnamefont {M.}~\bibnamefont
  {Li}}, \bibinfo {author} {\bibfnamefont {L.}~\bibnamefont {Chen}},\ and\
  \bibinfo {author} {\bibfnamefont {C.}~\bibnamefont {Chen}},\ }\href
  {https://doi.org/10.1103/PhysRevA.59.3109} {\bibfield  {journal} {\bibinfo
  {journal} {Phys. Rev. A}\ }\textbf {\bibinfo {volume} {59}},\ \bibinfo
  {pages} {3109} (\bibinfo {year} {1999})}\BibitemShut {NoStop}%
\bibitem [{\citenamefont {Faruk}\ and\ \citenamefont
  {Bhuiyan}(2015)}]{Faruk_2015}%
  \BibitemOpen
  \bibfield  {author} {\bibinfo {author} {\bibfnamefont {M.~M.}\ \bibnamefont
  {Faruk}}\ and\ \bibinfo {author} {\bibfnamefont {G.~M.}\ \bibnamefont
  {Bhuiyan}},\ }\href {https://doi.org/10.5506/APhysPolB.46.2419} {\bibfield
  {journal} {\bibinfo  {journal} {Acta Phys. Pol. B}\ }\textbf {\bibinfo
  {volume} {46}},\ \bibinfo {pages} {2419} (\bibinfo {year}
  {2015})}\BibitemShut {NoStop}%
\bibitem [{\citenamefont {Mukherjee}\ \emph {et~al.}(2019)\citenamefont
  {Mukherjee}, \citenamefont {Patel}, \citenamefont {Yan}, \citenamefont
  {Fletcher}, \citenamefont {Struck},\ and\ \citenamefont
  {Zwierlein}}]{mukherjee_2019_2}%
  \BibitemOpen
  \bibfield  {author} {\bibinfo {author} {\bibfnamefont {B.}~\bibnamefont
  {Mukherjee}}, \bibinfo {author} {\bibfnamefont {P.~B.}\ \bibnamefont
  {Patel}}, \bibinfo {author} {\bibfnamefont {Z.}~\bibnamefont {Yan}}, \bibinfo
  {author} {\bibfnamefont {R.~J.}\ \bibnamefont {Fletcher}}, \bibinfo {author}
  {\bibfnamefont {J.}~\bibnamefont {Struck}},\ and\ \bibinfo {author}
  {\bibfnamefont {M.~W.}\ \bibnamefont {Zwierlein}},\ }\href
  {https://doi.org/10.1103/PhysRevLett.122.203402} {\bibfield  {journal}
  {\bibinfo  {journal} {Phys. Rev. Lett.}\ }\textbf {\bibinfo {volume} {122}},\
  \bibinfo {pages} {203402} (\bibinfo {year} {2019})}\BibitemShut {NoStop}%
\bibitem [{\citenamefont {Yan}\ \emph {et~al.}(2022)\citenamefont {Yan},
  \citenamefont {Patel}, \citenamefont {Mukherjee}, \citenamefont {Vale},
  \citenamefont {Fletcher},\ and\ \citenamefont {Zwierlein}}]{yan_2022_2}%
  \BibitemOpen
  \bibfield  {author} {\bibinfo {author} {\bibfnamefont {Z.}~\bibnamefont
  {Yan}}, \bibinfo {author} {\bibfnamefont {P.~B.}\ \bibnamefont {Patel}},
  \bibinfo {author} {\bibfnamefont {B.}~\bibnamefont {Mukherjee}}, \bibinfo
  {author} {\bibfnamefont {C.~J.}\ \bibnamefont {Vale}}, \bibinfo {author}
  {\bibfnamefont {R.~J.}\ \bibnamefont {Fletcher}},\ and\ \bibinfo {author}
  {\bibfnamefont {M.}~\bibnamefont {Zwierlein}},\ }\href@noop {} {} (\bibinfo
  {year} {2022}),\ \Eprint {https://arxiv.org/abs/2212.13752}
  {arXiv:2212.13752} \BibitemShut {NoStop}%
\bibitem [{\citenamefont {Lysebo}\ and\ \citenamefont
  {Veseth}(2009)}]{Lysebo_2009_2}%
  \BibitemOpen
  \bibfield  {author} {\bibinfo {author} {\bibfnamefont {M.}~\bibnamefont
  {Lysebo}}\ and\ \bibinfo {author} {\bibfnamefont {L.}~\bibnamefont
  {Veseth}},\ }\href {https://doi.org/10.1103/PhysRevA.79.062704} {\bibfield
  {journal} {\bibinfo  {journal} {Phys. Rev. A}\ }\textbf {\bibinfo {volume}
  {79}},\ \bibinfo {pages} {062704} (\bibinfo {year} {2009})}\BibitemShut
  {NoStop}%
\bibitem [{\citenamefont {Kokkelmans}()}]{kokkelmans}%
  \BibitemOpen
  \bibfield  {author} {\bibinfo {author} {\bibfnamefont {S.~J. J. M.~F.}\
  \bibnamefont {Kokkelmans}},\ }\href@noop {} {}\bibinfo {howpublished}
  {private communication}\BibitemShut {NoStop}%
\end{thebibliography}
\end{document}